\documentclass{imsart}

\RequirePackage{amsthm,amsmath,amsfonts,amssymb,bm}
\RequirePackage[authoryear]{natbib}
\usepackage{tikz}
\usepackage{float}
\usepackage{svg}
\usepackage{url}
\usepackage{pifont}
\usepackage[margin=1in]{geometry}
\usetikzlibrary{positioning, calc, shapes.geometric, shapes, shapes.multipart, arrows.meta, arrows, decorations.markings, external, trees}
\RequirePackage{graphicx}

\startlocaldefs
\newtheorem{assumption}{Assumption}
\newtheorem{proposition}{Proposition}
\newtheorem{remark}{Remark}

\def \E{\mathbb{E}}
\def\underbar#1{\underline{\sbox\tw@{#1}\dp\tw@\z@\box\tw@}}
\def \bmu{\boldsymbol{\mu}_i}
\def \bdelta{\boldsymbol{\bar{\delta}}_t}
\def \blambda{\boldsymbol{\lambda}_t}
\def \bLambda{\boldsymbol{\Lambda}_t}
\def \bY{\boldsymbol{\bar{Y}}_{it}}
\def \P{\mathbb{P}}
\def \pto{\stackrel{p}\to}
\newcommand{\dbY}{\ensuremath{\ddot{\overline{\bm Y}}_{it}}}
\newcommand{\dbYn}{\ensuremath{\ddot{\overline{\bm Y}}}}
\def \bepsilon{\boldsymbol{\bar{\epsilon}}_{it}}
\def \bSigma{\boldsymbol{\bar{\Sigma}}_{it}}

\tikzstyle{Arrow} = [
	thick, 
	decoration={
		markings,
		mark=at position 1 with {
			\arrow[thick]{latex}
			}
		}, 
	shorten >= 3pt, preaction = {decorate}
	]

\endlocaldefs

\begin{document}

\begin{frontmatter}
\title{Autoregressive models for panel data causal inference with application to state-level opioid policies}

\begin{aug}
\author[A]{\fnms{Joseph} \snm{Antonelli}},
\author[B]{\fnms{Max} \snm{Rubinstein}},
\author[B]{\fnms{Denis} \snm{Agniel}},
\author[B]{\fnms{Rosanna} \snm{Smart}},
\author[C]{\fnms{Elizabeth A.} \snm{Stuart}},
\author[D]{\fnms{Matthew} \snm{Cefalu}},
\author[B]{\fnms{Terry} \snm{Schell}},
\author[B]{\fnms{Joshua} \snm{Eagan}},
\author[E]{\fnms{Elizabeth} \snm{Stone}},
\author[B]{\fnms{Max} \snm{Griswold}},
\and
\author[B]{\fnms{Beth Ann} \snm{Griffin}}

\address[A]{University of Florida, Department of Statistics, 102 Griffin-Floyd Hall, P.O. Box 118545, Gainesville, FL 32611}
\address[B]{RAND Corporation, 1776 Main St, Santa Monica, CA 90401}
\address[C]{Johns Hopkins Bloomberg School of Public Health, Department of Biostatistics, 615 N. Wolfe Street in Baltimore, Maryland 21205}
\address[D]{Disney Streaming, Los Angeles, California}
\address[E]{Rutgers Robert Wood Johnson Medical School, Department of Psychiatry,125 Paterson Street, New Brunswick, NJ 08901}

\end{aug}

\begin{abstract}
Motivated by the study of state opioid policies, we propose a novel approach that uses autoregressive models for causal effect estimation in settings with panel data and staggered treatment adoption. Specifically, we seek to estimate the impact of key opioid-related policies by quantifying the effects of must access prescription drug monitoring programs (PDMPs), naloxone access laws (NALs), and medical marijuana laws on opioid prescribing. Existing methods, such as differences-in-differences and synthetic controls, are challenging to apply in these types of dynamic policy landscapes where multiple policies are implemented over time and sample sizes are small. Autoregressive models are an alternative strategy that have been used to estimate policy effects in similar settings, but until this paper have lacked formal justification. We outline a set of assumptions that tie these models to causal effects, and we study biases of estimates based on this approach when key causal assumptions are violated. In a set of simulation studies that mirror the structure of our application, we show that our proposed estimators frequently outperform existing estimators. In short, we justify the use of autoregressive models to evaluate the effectiveness of four state policies in combating the opioid crisis.
\end{abstract}


\end{frontmatter}

\section{Introduction}
The U.S. continues to confront an opioid epidemic \citep{volkow2017role}. Between 1999 and 2019, nearly 500,000 people died from an overdose involving opioids (\cite{wonder}). The number of fatal overdoses increased nearly 30\% from 2019 to 2020, reflecting the spread of fentanyl and greater opioid misuse associated with the increased stress, social isolation, and job loss stemming from the COVID-19 pandemic \citep{cdc2020overdose,patel2021opioid}. 
In response, states continue to implement an array of policies meant to combat the opioid-crisis, producing a policy landscape that is both complex and dynamic \citep{davis2017state,davis2021laws,davis2020opioid,davis2021continuing,haffajee2018policy,burris2017state,smart2022legal,acep2022opioid,pacula2020state,andraka2022laws,andraka2022national,frank2021policy}. Unfortunately, it is unclear whether these policies have been effective, as existing research has produced divergent effect estimates of varying credibility. Producing credible effect estimates of these policies is crucial as policymakers choose strategies to address this crisis. In this paper, we therefore estimate the effects of four policies on opioid prescriptions: two types of naloxone access laws (NALs), must access prescription drug monitoring programs (PDMPs), and medical marijuana laws. 

Unfortunately, obtaining credible estimates of these effects is challenging. The statistical literature on estimating causal effects in panel data settings is extensive, and many existing methods are -- broadly speaking -- implementations and extensions of difference-in-differences (DiD) designs \citep{ashenfelter1984using, angrist2008mostly, callaway2021difference, goodman2021difference, ben2021augmented, ben2022synthetic} and synthetic control methods \citep{abadie2010synthetic, abadie2015comparative}. Despite many methodological advancements in these designs, the methods are not generally well suited to the policy landscape seen in the opioid policy evaluation, where multiple policies are implemented simultaneously, and disentangling the effects of each policy is important \citep{griffin2022cooccur}. While recent approaches have proposed extensions to synthetic controls or DiD to incorporate multiple policies \citep{fricke2017identification, agarwal2020synthetic, roller2023differences}, these methods are only applicable when only one of the multiple treatments is of interest. Additionally, even in settings with a single policy, existing estimators have limited statistical power \citep{griffin2021moving,griffin2023confounding}.

To address these shortcomings, \cite{schell2018evaluating} and \cite{griffin2021moving} proposed estimating effects using so-called ``debiased autoregressive models,'' which extend standard autoregressive models to estimate causal effects. The challenge with these standard models is that they control for lags of the dependent variable: in an analysis of policy effects, this will induce ``collider bias'' due to the fact that the lagged outcomes are likely affected by prior treatments \citep{achen2000lagged,elwert2014endogenous,pearl2009causality,nickell1981biases}. At a high level, debiased models address this problem by removing policy effects from prior outcomes in the estimation process. While lacking formal justification, a number of simulation studies showed that these models performed well relative to state-of-the-art implementations of difference-in-differences and synthetic controls when evaluating policy effects (\cite{schell2018evaluating}). Other studies demonstrated good performance under different confounding mechanisms, the presence of co-occurring policies, and treatment effect heterogeneity \citep{griffin2022cooccur,griffin2023confounding,griswold2023timevary}. 

Our primary methodological contribution is therefore to propose a novel causal identification result that justifies the use of debiased autoregressive models for estimating causal effects. Additionally, we study biases of estimates based on this approach when the causal assumptions are violated, particularly with respect to treatment effect heterogeneity. We find that the different estimators in this setting present a standard bias-variance trade-off and that debiased autoregressive models make a very favorable trade-off. To be precise, we show that bias is typically low for these estimators, and they can have significantly lower variance than existing estimators. This is critically important for the analysis of state opioid policy data, where both sample sizes and policy effect sizes may be small. Additionally, we formalize the assumptions required to use these models to estimate effects of multiple policies simultaneously, a challenging and relevant problem in opioid policy research. While we primarily seek to demonstrate that this method is useful to quantify the impact of opioid policies, these models are likely useful in both other policy evaluations and more general settings. 

Our paper proceeds as follows. In Section \ref{sec:DataDescription}, we introduce the state-level opioid data and the corresponding data structure for our application. In Section \ref{sec:methods}, we outline a simplified causal framework that motivates using debiased autoregressive models to estimate causal effects of a single policy. In Section \ref{sec:relaxations}, we discuss several extensions that relax several of the assumptions from Section \ref{sec:methods}, including estimating the effects of multiple policies jointly, and a discussion of the potential for bias in the presence of effect heterogeneity. In Section \ref{sec:Connections}, we discuss connections between our approach and other commonly used approaches, including difference-in-differences, synthetic controls, and methods based on sequential randomization \citep{robins2000marginal}. In Section \ref{sec:Simulations}, we provide a set of simulation studies that further supports the use of our method for our application and highlights when this approach may perform well. In Section \ref{sec:Application}, we estimate the effects of state policies on opioid prescriptions and compare our estimates against other existing approaches. Section \ref{sec:Discussion} concludes with further discussion of our methodological and applied contributions.

\section{State-level opioid policy data} \label{sec:DataDescription}

Our work is motivated by the need to assess the impact of four state-level policies, using data on the number of opioid prescriptions per capita in each state over the years 2006-2017. In this section, we provide background on the data and its structure. 

\subsection{Policies, outcomes, and covariates}

We obtained our policy measures from the OPTIC-Vetted Policy Data Sets from the RAND-USC Opioid Policy Tools and Information Center (OPTIC) \citep{randoptic2024mm,randoptic2024pdmp,randoptic2024naloxone}. We study four policies that have been examined in prior research but are rarely considered jointly: two types of naloxone access laws (NALs), must access prescription drug monitoring programs, and medical marijuana laws. 

Naloxone is a potential life saving treatment for opioid overdoses that works by reversing the effects of opioids \citep{cawley2023harm,chamberlain1994comprehensive}. However, naloxone may be difficult to obtain due to stigma or lack of access to traditional healthcare systems \citep{green2020laws,smart2021effectiveness}. To address these barriers, every state has adopted some type of NAL by 2018 \citep{smart2021systematic}. There are two broad classes of NALs \citep{NBERw31142}: those which allow distribution through a standing or protocol order or those which allow distribution through a pharmacist prescriptive authority \citep{davis2015legal,davis2017state}. We therefore consider each class as a separate policy. PDMPs are electronic databases that track prescribed and dispensed controlled substances within a state \citep{Schuler2020}. By giving healthcare providers information about a patient's prescription drug history, PDMPs may help identify patients who are at risk for opioid misuse or overdose, potentially increasing early intervention and support \citep{NBERw24947}. Based on the literature on PDMPs, state laws that specify that prescribers must access the PDMP before writing a prescription to a new/existing patient tends to allow for more effective forms of PDMPs \citep{pacula2020state,randoptic2024pdmp}. Thus, we define our PDMP laws as those which have this must access component. Finally,  medical marijuana laws legalize marijuana for medical purposes \citep{pacula2015assessing}. These laws may reduce opioid use and opioid-related harms by providing an alternative treatment option for chronic pain and other conditions that are commonly treated with opioids \citep{POWELL201829}.  For a summary of the prevalance of these laws over time, see Figure \ref{fig:PolicyWheel} below.\footnote{For all four policies, our policy measure is the fraction of the year each state has such a law in place.}

\begin{figure}[H]
    \centering
        \includegraphics[width=\textwidth]{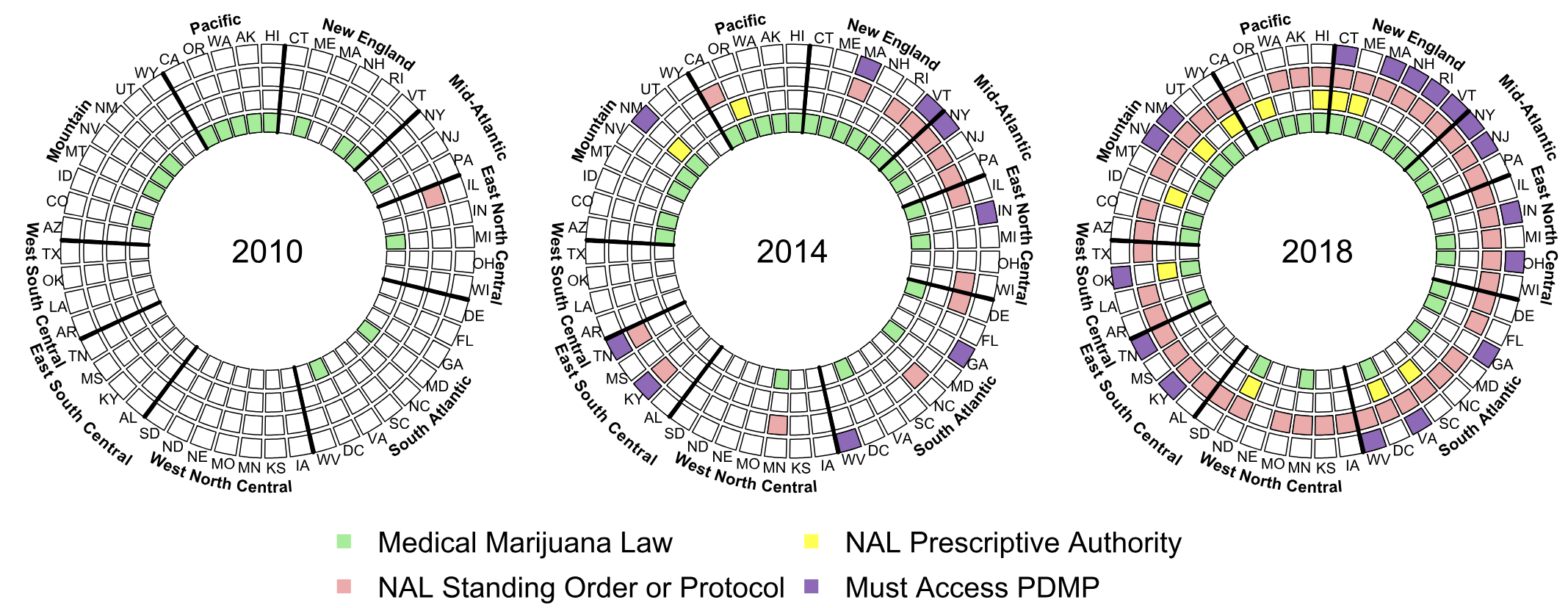}
    \caption{Policy wheels showing the timeline of policy adoption}
    \label{fig:PolicyWheel}
\end{figure}

\noindent We examined the impact of these policies on annual CDC opioid prescribing rates per 100,000 people from 2006-2017. In contrast to the policy measures, we construct the outcomes from the QuintilesIMS Transactional Data Warehouse (TDW). These data represent approximately 59,000 retail (non-hospital) pharmacies, which dispense nearly 88\% of all retail prescriptions in the U.S. We identify opioid prescriptions using the National Drug Code. We do not include cough and cold formulations, which contain opioids, or buprenorphine, an opioid partial agonist used for treatment of opioid use disorder and pain. To convert the prescription counts into rates, we use annual resident population estimates obtained from the Population Estimates Program, U.S. Census Bureau. An illustration of opioid prescription rates over time by state can be found in Appendix \ref{app:E}. 

Finally, we include seven time-varying covariates in our models. These are: annual unemployment rates from the St. Louis Federal Reserve’s FRED Economic Data system, median income and poverty rate derived from the U.S. Census Bureau’s 1-year American Community Survey (ACS), and percentages of non-Hispanic white individuals, females, individuals under 20 years old, and those over 64 years old, derived from the U.S. Census Intercensal (2000–2010) and Postcensal (2011–2022) estimates. These covariates were selected to capture a range of sociodemographic factors that might confound the effects of opioid policies on the outcomes.

\subsection{Data structure and summary}

Our data include 51 units (50 states plus Washington D.C.) observed over 12 years at yearly intervals. An illustration of the structure of our data for one of the policy variables is given in Table \ref{tab:DataStructurePolicy}, which illustrates that the policy variables follow a staggered adoption regime, where states adopt policies at different times, and policy adoption does not reverse during the course of our study. For more information on all of the variables in our data, Table \ref{tab:DataSummary} displays summary statistics for each variable in both the first and last year of our study. Opioid prescription rates drop over the course of our study. Additionally, at the beginning of the study, only medical marijuana laws were enacted among all four policies, whereas by the end many states had enacted these policies. Covariate values follow expected trends over time, such as increasing income over time. 

\begin{table}[ht]
\centering
\begin{tabular}{rrrrrrrrr}
  \hline
 & 2006 & 2007 & 2008 & 2009 & 2010 & \dots & 2016 & 2017 \\ 
  \hline
State 1 & \ding{55} & \ding{55} & \ding{55} & \ding{55} & \ding{55} & \dots & \ding{55} & \ding{55} \\ 
 State  2 & \ding{55} & \ding{55} & \ding{51} & \ding{51} & \ding{51} & \dots & \ding{51} & \ding{51} \\ 
 \vdots & \vdots & \vdots & \vdots & \vdots & \vdots & \vdots & \vdots & \vdots \\ 
 State  50 & \ding{55} & \ding{55} & \ding{55} & \ding{55} & \ding{55} & \dots & \ding{51} & \ding{51} \\ 
 State  51 & \ding{55} & \ding{55} & \ding{55} & \ding{55} & \ding{51} & \dots & \ding{51} & \ding{51} \\ 
   \hline
\end{tabular}
\caption{Hypothetical illustration of data structure for a single policy. Checks indicate a year in which the policy was enacted. Note there are 51 states as we include Washington D.C as a unit of analysis. }
\label{tab:DataStructurePolicy}
\end{table}

\begin{table}[ht]
\centering
\begin{tabular}{lrll}
  \hline
& & Year 2006 & Year 2017 \\ 
  \hline
  \hline
Covariates & Median income (\$) & 24154 (3239) & 30407 (4469) \\ 
&   Unemployment rate (\%) & 4.48 (1.08) & 4.16 (0.93) \\ 
&   Percentage White (\%) & 72.43 (16.10) & 68.26 (16.16) \\ 
&   Percentage female (\%) & 50.74 (0.79) & 50.59 (0.84) \\ 
&  Percentage less than 20 (\%) & 27.33 (1.90) & 25.15 (2.07) \\ 
&   Percentage above 64 (\%) & 12.65 (1.67) & 15.96 (1.91) \\ 
&   Percentage in poverty (\%) & 12.67 (3.07) & 12.78 (2.82) \\ 
\hline
\hline
Policies &   Naloxone standing (\%) & 0 (0) & 85.29 (32.13) \\ 
&   Naloxone prescriptive (\%) & 0 (0) & 15.69 (34.55) \\ 
&   PDMP must access laws (\%) & 0 (0) & 33.33 (47.61) \\ 
&   Medical marijuana (\%) & 23.53 (42.84) & 56.21 (49.66) \\ 
\hline
\hline
Outcome & Prescription rate (per 100,000) & 76.71 (21.69) & 62.59 (17.91) \\ 
   \hline
\end{tabular}
\caption{Mean and standard deviation of variables in our study in 2006 and 2017}
\label{tab:DataSummary}
\end{table}

\section{Methods} \label{sec:methods}

We now motivate our methodological approach, debiased autoregressive models, and first introduce a simplified version of our causal identification strategy, which suggests a simple parametric modeling approach to estimation. While facilitating exposition, this simplicity comes at the expense of strong identifying assumptions and restrictive model parameterizations. We defer outlining relaxations of these assumptions to Section \ref{sec:relaxations}, though we rely on these relaxations for our analysis of state-level opioid data in Section \ref{sec:Application}. We begin by formalizing the causal estimands that we wish to estimate, and then outline the set of simplified assumptions that allow us to identify these estimands in terms of the observed data distribution. 

\subsection{Simplified causal framework}

\subsubsection{Estimands}

We consider a setting where we observe $Y_{it}$,  an outcome of the observation $i$ at time $t$, with $i=1,\dots,n$ and $t=0,\dots,T$. We let $\bar{Z}_{it}$ denote the history of some random variable $Z$ from time $t = 0, \dots, t$. We also let $A_{it}$ denote an indicator of a single policy of interest being active for observation $i$ at time $t$, and define $Y_{it}(\bar{a})$ as the potential outcome that would be observed if the treatment history $\bar{A}_{it}$ was fixed at the level $\bar{a}$. Finally, we let $T_{0i} + 1$ denote the time that unit $i$ first adopts policy $A_{it}$. In contrast to our real data structure, we do not yet consider other policies or covariates. Our goal is to estimate the average treatment effect across all study time periods, given by 
\begin{align*}
\theta = \mathbb{E}[Y_{it}(1) - Y_{it}(0)]. 
\end{align*}
\noindent This estimand makes no restrictions on heterogeneity over time or across units, though averages over both. Throughout this section and Section \ref{sec:relaxations}, we focus on this population level estimand as it easily allows us to study identification and biases under violations of our assumptions. At times in the manuscript, however, we instead consider a sample-level average treatment effect on the treated (SATT),
\begin{align*}
\theta^\mathcal S_m=\frac{1}{n_t} \sum_{i \in \mathcal{N}_t} \sum_{m=1}^{M} \bigg\{ Y_{i, T_{0i} + m}(1) - Y_{i, T_{0i} + m}(0) \bigg\},
\end{align*}

\noindent where $\mathcal{N}_t$ contains the indices of treated units, and $n_t$ is the number of treated units. This estimand reflects the average treatment effect on the treated units in the first $M$ time periods after policy adoption. Our proposed framework allows for estimation of either quantity, though we leave details on performing inference on $\theta^\mathcal S_m$ to Appendix C. Many popular methods, such as difference-in-differences and synthetic controls, target $\theta^\mathcal S_m$ in the staggered adoption setting. We therefore consider this quantity in order to compare our approach to other existing approaches in simulations and for the opioid policy application. However, we also consider more complex estimands that examine multiple policies simultaneously or depend on prior treatment values in Section \ref{sec:relaxations}.

A major challenge when estimating these causal effects is that we never observe the outcomes when $A_{it} = 1$ and $A_{it} = 0$ for any individuals. We therefore must make assumptions tying both potential outcomes $Y_{it}(1)$ and $Y_{it}(0)$ to the observed data distribution. We discuss a set of assumptions sufficient to identify $\theta$ in terms of the observed data distribution below.

\subsubsection{Assumptions}

We again first introduce a simple set of assumptions that are sufficient to identify $\theta$ in terms of the observed data distribution, and we defer relaxing these assumptions to Section \ref{sec:relaxations}. We first make the consistency assumption,

\begin{assumption}[Consistency]\label{asmpt-consistency}
$\bar{A}_{it} = \bar{a} \implies Y_{it} = Y_{it}(\bar{a}).$
\end{assumption}

\noindent Consistency states that the observed outcome under the observed policy history $\bar{A}_{it} = \bar{a}$ is equal to the potential outcome under that same policy history. This would be violated if there were multiple versions of the policy or when observed outcomes depend on other units' policies, sometimes referred to as ``interference'' or ``spillover effects.'' 

We next assume no carryover effects,

\begin{assumption}[No carryover effects]\label{asmpt-nce}
For any treatment history $\bar{a}$, $Y_{it}(\bar{a}) = Y_{it}(a_{t})$.
\end{assumption}

\noindent No carryover effects limits the dependence of the potential outcomes on the treatment history to only the current treatment status. In other words, we assume that prior treatments have no impact on the current outcomes. While strong, this assumption places no restrictions on how treatment effects may vary over time, and is common in the policy evaluation literature. 

We next assume ignorability, which states that treatment assignment $A_{it}$ is effectively randomized with respect to the potential outcomes in time period $t$ given $Y_{i,t-1}(0)$, the previous potential outcome absent treatment, which we will call the \textit{prior control counterfactual}.

\begin{assumption}[Ignorability]\label{asmpt-ignorability}
$Y_{it}(a) \perp A_{it} \mid Y_{i,t-1}(0), \quad a=0, 1.$
\end{assumption}

\noindent In contrast to standard ignorability assumptions, we assume that randomization occurs with respect to a potentially unobserved quantity, the prior control counterfactual, which is unobserved for those who had already been treated. We provide a more complete discussion of this identifying assumption in comparison to other assumptions in Sections \ref{ssec:did} and \ref{ssec:gfm}. 

We next assume treatment effects are constant across units:

\begin{assumption}[Constant treatment effects]\label{asmpt-uniform}
$Y_{it}(a) = Y_{it}(0) + a \theta, \quad a = 0, 1.$
\end{assumption}

\noindent Constant treatment effects implies that the difference between the two counterfactuals is a fixed quantity across units and time (i.e. $Y_{it}(1) - Y_{it}(0) = \theta$) and it precludes any kind of random heterogeneity. This assumption is especially strong, particularly if certain units are expected to respond to treatment differently. 

We also assume positivity,

\begin{assumption}[Positivity]\label{asmpt-positivity}
For some $\epsilon > 0$ and $e(y) = P\left\{A_{it} = 1 \mid Y_{i,t-1}(0) = y\right\}$, 
\begin{align*}
P\left[\epsilon < e\{Y_{i,t-1}(0)\} < 1 -\epsilon\right] = 1.
\end{align*}
\end{assumption}

\noindent This requires that we have positive probability of observing either policy status conditional on the prior counterfactual under control. 

Finally, we assume that the expected potential outcomes follow some known parametric model.

\begin{assumption}[Known parametric model]\label{asmpt-linear}
Assume that for each unit $i$ and time period $t$
\begin{align}
\label{eqnres1}Y_{it}(0) = \alpha_t +\beta_t^\top\phi(Y_{i,t-1}(0)) + \epsilon_{it},
\end{align}
\noindent for some parameter vector $(\alpha_1, \beta_1^\top, \dots, \alpha_T, \beta_T^\top)^\top \in \mathbb{R}^{T(q+1)}$ and some known function $\phi(x): \mathbb{R}\to\mathbb{R}^q$. Let $Z_{it} = [1, A_{it}, \phi^\top(Y_{it-1}(0))]$ and assume that $\E[Z_{it}^\top Z_{it}]$ has full column rank.
\end{assumption}

\begin{proposition}\label{prop:para-ident}
Under assumptions \ref{asmpt-consistency}-\ref{asmpt-linear}, $\theta$ is identified in terms of the observed data distribution as the minimizer of $\E[(Y_{it} - \theta A_{it} - \alpha_t - \beta_t^\top\phi(Y_{i,t-1} - \theta A_{i, t-1}))^2]$.     
\end{proposition}

There is no general non-parametric identification result in this setting, so our proposed identification strategy relies on parametric models for identification. While the parametric assumptions may seem particularly strong, in practice one may choose a flexible set of functions $\phi$ to model the outcomes. Moreover, in many common scenarios $\theta$ may be identified non-parametrically. For example, there is often a time period $t^\star$ in which no units are treated, so that $Y_{i,t^\star}(0) = Y_{i,t^\star}$ for all units. In this case, Proposition \ref{prop:nonpara-ident} shows that non-parametric identification can be obtained.

\begin{proposition}\label{prop:nonpara-ident}
    Assume (\ref{asmpt-consistency})-(\ref{asmpt-positivity}), and that there exists some time period $t^\star$ where $A_{it^\star} = 0$ for all units $i$. Then,
    \begin{align}\label{eqn:sufficient}
    \theta = \E[\E[Y_{i,t^\star+1} \mid A_{i,t^\star+1} = 1, Y_{it^\star}] - \E[Y_{i,t^\star+1} \mid A_{i,t^\star+1} = 0, Y_{it^\star}]].
    \end{align}
\end{proposition}

\noindent Equation (\ref{eqn:sufficient}) in Proposition \ref{prop:nonpara-ident} shows that $\theta$ can be identified in the data by modeling the conditional expectation function using outcomes from the $t^\star+1$ time period. Additionally, in some cases we may want to assume that $\theta$ varies over time -- i.e. we may assume that equation \eqref{eqn:sufficient} only identifies $\theta_{t^\star+1}$. In this case, we may recursively identify $\theta_t$ for any time period $t > t^\star + 1$ using

\begin{align*}
\theta_t = \E[\E[Y_{it} \mid A_{it} = 1, Y_{i,t-1} - A_{i,t-1} \theta_{t-1}] - \E[Y_{it} \mid A_{it} = 0, Y_{i,t-1} - A_{i,t-1} \theta_{t-1}]],
\end{align*}

\noindent suggesting a corresponding estimation procedure that iteratively estimates each $\theta_t$. Note that this non-parametric identification strategy only uses data from a single time period to estimate the effect at that time, which leads to a reduction in efficiency. Because of this, we focus on our parametric modeling approach, which borrows information across time periods by assuming a constant treatment effect across time, or by assuming it can be modeled as a parametric function of time as we will see in subsequent sections. 

\subsection{Estimation}
Under Assumptions \ref{asmpt-consistency}-\ref{asmpt-linear}, we specify a debiased linear autoregression model (L-DAM) for the conditional expectation of the outcome \citep{benjamin2003generalized, zeger1988markov}:

\begin{align}\label{eqn:ldam}
    Y_{it} &= \alpha + \beta (Y_{i,t-1} - \theta A_{i,t-1}) + \theta A_{it} + \epsilon_{it} 
\end{align}

\noindent where $\E[\epsilon_{it} \mid A_{it}, Y_{i,t-1}(0)] = 0$, and we assume that the errors have zero correlation across units or over time (i.e. $Cov(\epsilon_{it}, \epsilon_{js}) = 0$ for $j\ne i, t \ne s$). The L-DAM reflects a simple linear parameterization of the assumptions above. We may use M-estimation to estimate these parameters and to conduct inference. Alternatively, we may specify priors for the parameters in \eqref{eqn:ldam} and use Bayesian inference, the approach we adopt in the rest of the paper. We also suggest an IPW estimator and a doubly-robust approach that consistently estimates $\theta$ if either the IPW or outcome model are misspecified in Appendix \ref{app:dr}.

\section{Relaxing assumptions and extensions}\label{sec:relaxations}

We have focused on a simple set of identification assumptions to provide a straightforward and intuitive motivation for a simplified version of our proposed estimation strategy. While useful for exposition, these assumptions were quite strong and may be weakened substantially. We now consider both relaxations of these assumptions, and how violations of the assumptions may bias our estimates. Specifically, for this latter piece we focus our discussion on Assumption 4, which enforces constant treatment effects. While ignoring treatment effect heterogeneity can negatively impact results in many ways that are not unique to our setting, the particular form of misspecification we study is unique to our identification strategy, and we therefore focus much of the discussion on this problem, including a mathematical argument that ignoring this heterogeneity will not generally induce large biases. Finally, we consider extensions that relax other assumptions, such as allowing for additional time dependence, more complex estimands allowing for different treatment sequences, time-varying covariates, multiple policies, and non-linear outcome models. 

\subsection{Effect heterogeneity}\label{sec:Heterogeneity}

Assumption \ref{asmpt-uniform} is an especially strong assumption required for identification and estimation. Under our modeling assumptions, it requires that $Y_{i,t-1} - \theta A_{i,t-1}$ is equal to $Y_{i, t-1}(0)$. In the presence of treatment effect heterogeneity, this equivalence will not hold, leading standard estimation strategies to have both biased effect and variance estimates. Moreover, as heterogeneity increases, the degree of this bias increases. We first discuss a simple solution to this problem: modeling the effect heterogeneity as a function of observed covariates. While a simple solution, we argue that in many applications this won't be necessary, and later use mathematical derivations to argue that the bias due to effect heterogeneity is likely small in many settings.

\subsubsection{Modeling the effect heterogeneity}

A simple approach to allow effect heterogeneity is to model it within the autoregressive framework. For example, if $\E[Y_{it}(1)-Y_{it}(0) \mid V_{it}] = h(V_{it}, t)$, where $V_{it}$ includes some subset of the observed covariates $X_{it}$, we can specify a parametric form for $h(\cdot)$ and adjust for $Y_{i,t-1} - A_{i,t-1} h(V_{i,t-1}, t-1)$.\footnote{It is also potentially feasible to allow heterogeneity with the prior period control counterfactuals. However, this would lead to a recursive formula requiring a time period where no one was treated in order to estimate the equation.} A well-specified $h(\cdot)$ function will eliminate effect heterogeneity induced biases and allow us to study whether certain characteristics modify the treatment effect. However, one practical issue, particularly when analyzing state-level data, is that we already have little data to estimate treatment effects. This additional modeling may lead to overly variable effect estimates, as well as reduce power to estimate average effects. 

We therefore propose an approach to estimating heterogeneous treatment effects through a parametric model for $h(V_{it}, t)$ that can simultaneously identify treatment effect heterogeneity when it is present, while also reducing the impact of estimation uncertainty when the effects are at least approximately homogeneous or we have insufficient data to fit such models. Specifically, we assume the additive model
\begin{align*}
h(V_{it}, t) = \nu_0 + b(t) + \sum_{j} V_{it,j} \nu_j,
\end{align*}
\noindent where $b(\cdot)$ can be fit using any one-dimensional smoothing approach, though we focus our attention on estimation of the $\nu_j$ coefficients and heterogeneity by covariates rather than heterogeneity over time. Note that while we have included covariates $V_{it}$ into this model linearly, nonlinear functions of these variables could also be incorporated in a similar manner. Placing non-informative prior distributions on the $\nu_j$ coefficients might be the preferred strategy if we had large amounts of data; however,  this will lead to overly variable estimates in small data settings or when the number of covariates included in $V_{it}$ is large. To reflect the fact that treatment effects both 1) may indeed be homogeneous, or at least approximately so, and 2) we may not have enough data to estimate these effects, we incorporate recent prior distributions in the high-dimensional Bayesian statistics literature. We can use standard, non-informative Gaussian prior distributions for $\nu_0$ and the coefficients of $b(\cdot)$ as we do not want to shrink the overall treatment effect towards zero. For the $\nu_j$ parameters that dictate heterogeneity by observed covariates or prior counterfactuals, we can use the horseshoe prior distribution \citep{carvalho2009handling} to shrink this heterogeneous treatment effect function towards a homogeneous effect. This prior distribution takes the form
\begin{align*}
    \nu_j &\sim \mathcal{N}(0, \upsilon  \omega_j) \\
    \upsilon &\sim C^+(0, 1) \\
    \omega_j &\sim C^+(0, 1) \quad \text{for all } j,
\end{align*}

\noindent where $C^+(0, 1)$ represents the half-Cauchy prior distribution. This prior distribution has been shown to aggressively shrink coefficients corresponding to non-important predictors towards zero, while reducing bias in the estimation of large coefficients. This allows us to accommodate heterogeneity without sacrificing too much efficiency, which is critical in state-level policy applications. Additionally, we can use the posterior distribution of $\boldsymbol{\nu}$ to test the null hypothesis that there is no heterogeneity present. This null hypothesis corresponds to $\nu_j = 0$ for all $j$, and we may construct simultaneous posterior credible intervals for all $\nu_j$ to address multiplicity concerns and examine whether zero is contained in the posterior distribution. Moreover, in panel data settings, there are potentially other tests one could construct that are not unique to the statistical model presented here. Roughly speaking, if treatment effect heterogeneity is present, we would expect more variability in treated units than control units. Similar ideas were explored in the context of randomized trials in \cite{mills2021detecting}. Extending these to the observational, panel data setting here is beyond the scope of this paper, but is a potential topic for future research. 

Of course, this strategy only works if the heterogeneity is a function of $V_{it}$. When it is not, this approach may still result in biased effect estimates for average causal effects. Fortunately, as we show formally below, in many instances it may be reasonable to assume that the heterogeneity, and therefore the induced bias, is small -- even when we do not model the heterogeneity. We also explore this in our simulation study in Section \ref{sec:Simulations}, where we show that heterogeneity of the treatment effect only impacts inference on the average treatment effect when the degree of heterogeneity is very large. Ignoring this heterogeneity may therefore often be a reasonable option in practice, particularly in situations where the average treatment effect is of primary substantive interest.  

\subsubsection{Effect heterogeneity as measurement error}

We now derive an expression for the bias due to effect heterogeneity when constant treatment effects are assumed. We do this by reframing this as a measurement error problem. Suppose Assumption \ref{asmpt-uniform} does not hold. Define $\theta_{it} = Y_{it}(1) - Y_{it}(0)$ as the unit-level treatment effect for unit $i$ at time $t$ such that $\theta = \mathbb{E}(\theta_{it})$. We can re-write \eqref{eqn:ldam} to obtain
\begin{align}
    Y_{it} &= \alpha + \beta(Y_{i,t-1}(0) + A_{i,t-1} \theta_{i,t-1} - A_{i,t-1} \theta) + \theta A_{it} + \epsilon_{it} \nonumber \\
    &= \alpha + \beta(Y_{i,t-1}(0) + U_{i,t-1}) + \theta A_{it} + \epsilon_{it} \label{eqn:YmodMeasurement}.
\end{align}
\noindent In other words, adjusting for $Y_{i,t-1} - A_{i,t-1} \theta$, our proxy of $Y_{i,t-1}(0)$, can be equivalently thought of as adjusting for a noisy version of $Y_{i,t-1}(0)$ subject to measurement error. An important parameter therefore becomes $\text{Var}(U_{i,t-1})$, which controls the magnitude of the heterogeneity and measurement error. Clearly, if $\text{Var}(U_{i,t-1}) = 0$ and we have no heterogeneity, then Assumption \ref{asmpt-uniform} holds, and $\theta$ is identified. 

Throughout the rest of this section, we focus on a simplified $\widehat{\theta}$ that comes from estimating the model in \eqref{eqn:YmodMeasurement}, ignoring the fact that $U_{i,t-1}$ is a function of $\theta$, and focusing on the impact of this sort of measurement error in this setting.\footnote{We find identical results when fitting the model in \eqref{eqn:ldam}, so we proceed with the simpler setting that has easier to understand mathematical derivations.} Measurement error in a confounder has been studied previously \citep{carroll1998measurement}, and it was shown that if $U_{i,t-1}$ is mutually independent of $Y_{i,t-1}(0), A_{it}$, and $Y_{it}$, then 
\begin{align}
    \mathbb{E}(\widehat{\theta}) = \theta + \beta \gamma_A \bigg( \frac{\text{Var}(U_{i,t-1})}{\text{Var}(U_{i,t-1}) + \sigma_{Y_{i,t-1}(0) \mid A_{it}}^2} \bigg), \label{eqn:ClassicalFormula}
\end{align}

\noindent where $\sigma_{Y_{i,t-1}(0) \mid A_{it}}^2$ is the residual variance from regressing $Y_{i,t-1}(0)$ against $A_{it}$, and $\gamma_A$ is the coefficient for $A_{it}$ in this regression. If we did not adjust for $Y_{i,t-1}(0)$ at all, we would have $\mathbb{E}(\widehat{\theta}) - \theta = \beta \gamma_A$. In other words, adjusting for an error-prone version of $Y_{i,t-1}(0)$ instead of $Y_{i,t-1}(0)$ itself biases estimates towards what we would obtain without adjusting for the confounder at all. Moreover, the degree of this bias depends on the relative size of $\text{Var}(U_{i,t-1})$ and $\sigma_{Y_{i,t-1}(0) \mid A_{it}}^2$.\footnote{In state-level studies of opioid policies, we expect the treatment effect $\theta$ to be relatively small in magnitude relative to the overall degree of variation in the outcomes, leading this bias to be small.} Proposition \ref{prop:1} formalizes reasoning that may be used to justify the proposed estimator. 

\begin{proposition}\label{prop:1}
    Under Assumptions \ref{asmpt-consistency}, \ref{asmpt-nce}, \ref{asmpt-ignorability}, \ref{asmpt-positivity} and assuming model \eqref{eqn:YmodMeasurement} holds, if $U_{i,t-1}$ is mutually independent of $Y_{i,t-1}(0), A_{it}$, and $Y_{it}$, and there exists a $k > 0$ such that $\frac{\text{Var}(U_{i,t-1})}{\sigma_{Y_{i,t-1}(0) \mid A_{it}}} < k$, then we have
    \begin{align*}
        \left|\mathbb{E}(\widehat{\theta}) - \theta\right| \leq |\beta \gamma_A| \frac{k}{k+1}
    \end{align*}
\end{proposition}

\noindent Proposition \ref{prop:1} shows that only large amounts of heterogeneity would lead to substantial bias in the estimation of the treatment effect when the measurement error follows a classical structure and is independent of the treatment, outcome and potential outcomes. However, the classical structure may not hold in many settings. Next, we discuss the impacts of differential measurement error.  

\subsubsection{Differential measurement error and sources of heterogeneity}

If the measurement error terms $U_{i,t-1}$ are correlated with any of $A_{it}$, $Y_{it}$, or $Y_{it}(0)$ then the bias expression in \eqref{eqn:ClassicalFormula} no longer holds. The exact bias expression in this setting can be found in Appendix B, where we discuss the impacts of differential measurement error from treatment effect heterogeneity in more detail. While we leave most details to the Appendix, we focus here on a particular form of differential measurement error that is most likely to occur and is most problematic. 

The most likely reason in our context for differential measurement error arises from correlation between $U_{i,t-1}$ and $Y_{it}$. When $U_{i,t-1}$ is high, unit $i$ was treated and had a large treatment effect at time $t-1$. When unit-level treatment effects are smooth over time, a large value of $U_{i,t-1}$ likely correlates with a high value of $Y_{it}$ due to the large unit-level treatment effect. Applying the formula from Appendix B to the situation where $U_{i,t-1}$ is correlated with $Y_{it}$, but not $A_{it}$, the estimate of the treatment effect has expectation

\begin{align*}
    \mathbb{E}(\widehat{\theta}) &= \frac{C_1 + C_2 + C_3}{\text{Var}(A_{it}) - \frac{\text{Cov}^2(A_{it},Y_{i,t-1}(0))}{\text{Var}(Y_{i,t-1}(0) + U_{i,t-1})}} \\
    &= \theta + \beta \gamma_A \bigg( \frac{\text{Var}(U_{i,t-1})}{\text{Var}(U_{i,t-1}) + \sigma_{Y_{i,t-1}(0) \mid A_{it}}^2} \bigg) + \frac{C_3}{\text{Var}(A_{it}) - \frac{\text{Cov}^2(A_{it},Y_{i,t-1}(0))}{\text{Var}(Y_{i,t-1}(0) + U_{i,t-1})}},
\end{align*}
where $C_1 = \text{Cov}(A_{it},Y_{it})$,
\begin{align*}
    C_2 = -\frac{ \text{Cov}(A_{it},Y_{it}(0))  \text{Cov}(Y_{it},Y_{it}(0))}{\text{Var}(Y_{i,t-1}(0) + U_{i,t-1})}, \quad
    C_3 = -\frac{ \text{Cov}(A_{it},Y_{it}(0))  \text{Cov}(Y_{it},U_{i,t-1})}{\text{Var}(Y_{i,t-1}(0) + U_{i,t-1})}.
\end{align*}

\noindent Notice that when the measurement error is classical and $\text{Cov}(Y_{it},U_{i,t-1}) = 0$, then $C_3 = 0$ and we revert back to the value in (\ref{eqn:ClassicalFormula}). 

We can examine whether $C_3$ is small (or large) in magnitude compared with $C_1$ and $C_2$ to assess how differential measurement error affects the  magnitude of bias. We can see that the ratio of $C_3$ and $C_2$ can be written as,
\begin{align*}
    \frac{\text{Cov}(Y_{it},U_{i,t-1})}{\text{Cov}(Y_{it},Y_{it}(0))} = \frac{\text{Corr}(Y_{it},U_{i,t-1}) \sqrt{\text{Var}(U_{i,t-1})}}{\text{Corr}(Y_{it},Y_{it}(0)) \sqrt{\text{Var}(Y_{it}(0))}}. 
\end{align*}
\noindent We typically expect $\text{Corr}(Y_{it},Y_{it}(0))$ to be high and likely larger in magnitude than $\text{Corr}(Y_{it},U_{i,t-1})$. Therefore, when $\text{Var}(U_{i,t-1})$ is much less than $\text{Var}(Y_{it}(0))$, differential measurement error will contribute little to the overall bias, and we will obtain bias similar to the expression given in Proposition \ref{prop:1}. Moreover, this bias is expected to be small absent large effect heterogeneity.

In summary, the mathematical expressions for the bias induced by effect heterogeneity when constant treatment effects are assumed suggest that for many practical applications, we may reasonably assume this bias to be small. 

\subsection{Additional time dependence}

We may also wish to relax the no carryover effects (assumption \ref{asmpt-nce}) and ignorability assumptions (assumption \ref{asmpt-ignorability}), allowing for greater dependence between the treatments and outcomes over time. First consider no carryover effects: in many policy settings, treatment in earlier time periods likely affects outcomes at time $t$, and therefore, the effects of entire treatment sequences are of interest. To incorporate these effects we first redefine our causal estimand, now given by $\theta_t(\bar{a}) = \mathbb{E}[Y_{it}(\bar{a}) - Y_{it}(\bar{0})].$ This parameter represents the effect of a particular treatment sequence compared against no treatment at time $t$. While in principle this allows for effects of all prior treatments, in practice, it is often useful to make the assumption that these ``carryover effects'' are limited to the most recent $l$ time points.

\begin{assumption}[$l$-limited carryover effects]\label{asmpt2}
Let $\bar{a}$ and $\bar{a}'$ be two treatment sequences such that $a_t = a_t'$, $a_{t-1} = a_{t-1}', \dots, a_{t-l} = a_{t-l}'$, then we have $Y_{it}(\bar{a}) = Y_{it}(\bar{a}'), \ \forall \bar{a},\bar{a}'.$
\end{assumption}

\noindent Assumption \ref{asmpt2} states that the potential outcomes for an entire treatment sequence only depend on the previous $l$ treatment values. Notice that when $l=0$ we return to the no carryover effects assumption. For simplicity, when we refer to $\bar{a}$ or $\bar{A}_{it}$ in what follows, we only refer to the treatment sequence for the previous $l$ time periods. 

Second, we relax Assumption \ref{asmpt-ignorability} to condition on $k$ prior time periods' counterfactuals as opposed to only the single time period's counterfactual outcome used previously. This relaxation increases the plausibility of the assumption, particularly in settings where we worry about the dependence of the outcomes over time. However, in order to relax this assumption, we must strengthen the positivity assumption (Assumption \ref{asmpt-positivity}) to include these additional time-periods. We may incorporate these relaxations into our estimation strategy by fitting the following version of the L-DAM:

\begin{align*}
    Y_{it} &= \alpha + \sum_{b=1}^k \left\{\beta_b \left(Y_{i,t-b} - \sum_{z=0}^l\theta_z A_{i,t-b-z} \right)\right\} + \sum_{z=0}^l \theta_z A_{i,t-z} + \epsilon_{it}.
\end{align*}

\noindent It is easy to see how this model incorporates these extensions. First, the $\theta_z$ parameters represent the lagged policy effects. Additionally, the model includes linear adjustments for estimates of $Y_{i,t-1}(0), \dots, Y_{i,t-k}(0)$ instead of estimates of only $Y_{i,t-1}(0)$.

\subsection{Time-varying covariates}\label{ssec:covariates}

We can also relax our ignorability assumption to allow for both time-invariant and time-varying covariates. Assume, for example, that covariates $X_{it}$ are determined prior to $Y_{it}$. Then we could assume that ignorability holds conditional on covariates and $k$ lagged potential outcomes absent treatment $(X_{it}$, $Y_{it-k}(0), ..., Y_{it-1}(0))$. However, including these covariates also requires a stronger positivity requirement to further condition on these quantities. One can then incorporate these same covariates into the L-DAM as:
\begin{align*}
    Y_{it} &= \alpha + \sum_{b=1}^k \left\{\beta_b \left(Y_{i,t-b} - \sum_{z=0}^l\theta_z A_{i,t-b-z} \right)\right\} + \sum_{z=0}^l \theta_z A_{i,t-z} + \Gamma^\top X_{it} + \epsilon_{it}.
\end{align*}

\noindent When the covariates $X_{it}$ are time-varying, causal identification also requires that these covariates are exogenous, or unaffected by treatment. That is, we must assume,

\begin{assumption}[Covariate exogeneity] $X_{it}(\bar{a}) = X_{it}(\bar{a}') = X_{it}, \qquad \forall \bar{a}, \bar{a}'.$
\end{assumption}

\noindent This assumption is frequently made in the applied policy literature, including when using two-way fixed effects models with time-varying covariates \citep{wooldridge2023simple}. In cases where this assumption is implausible, one can instead simply adjust for pre-treatment lags of the covariates, which, absent anticipatory effects, are unaffected by treatment by design.

\subsection{Multiple policies}

A separate but important type of time-varying variable we may also observe are other policy indicators $M_{it}$. For example, state legislatures may consider multiple policies to tackle opioid-related outcomes, including both naloxone access laws and Medicaid prescription coverage policies. In this situation, we may wish to know the effect of one policy, holding the other policy fixed at some value. In the language of potential outcomes, we are interested in a so-called ``controlled direct effect'' \citep{nguyen2021clarifying}.To be precise, assume that in each time period states may assign either policies $A$ and/or $M$, but we are primarily interested in the effects of policy $A$. One relevant causal quantity would then be $\psi_t(\bar{a}, \bar{m}) = \mathbb{E}[Y_{it}(\bar{a}, \bar{m}) - Y_{it}(\bar{0}, \bar{m})]$, where potential outcomes are now indexed by both treatment values. Identifying this quantity in terms of the observed data distribution requires analogues to Assumptions (\ref{asmpt-consistency})-(\ref{asmpt-positivity}). We formally describe these assumptions in Appendix A; however, intuitively we can view $(A, M)$ as a multi-valued treatment randomly assigned conditional on the prior potential outcomes absent any policy.\footnote{We can allow that the randomization for $M$ is conditional on $A$ or vice versa. In much of the mediation literature it is assumed that $A$ is determined prior to $M$. We instead make no assumption on their causal ordering, at the price of a slightly stronger conditional independence assumption.} Under these assumptions we can identify the causal estimand in terms of the observed data distribution as an average of a conditional expectation function, described in Appendix A. For continuous outcomes we can again parameterize a model for this function using an L-DAM: {\small
\begin{align*}
    Y_{it} &= \alpha +  \sum_{b=1}^k \left\{\beta_b \left(Y_{i,t-b} - \sum_{z=0}^l[\theta_z A_{i,t-b-z} + \zeta_z M_{i,t-b-z}] \right) \right\}
    + \sum_{z=0}^l [\theta_z A_{i,t-z} + \zeta_z M_{i,t-z}] + \epsilon_{it}
\end{align*}
}
\noindent Under this model, $\theta_z$ represents the direct effect of policy indicator $A_{t-z}$ on the outcome $Y_{it}$ holding the policy sequence $\bar{M}_{it}$ constant. As before, we can expand this model to include covariates and more complicated forms of treatment effects that allow, for example, effects due to interactions between policies $A$ and $M$. However, in small data settings, it may not be feasible to estimate these interactions with adequate precision.

\subsection{Nonlinear extensions}
\label{ssec:Nonlinear}

Linear models may be inappropriate for modeling the conditional expectation function for certain outcomes. For example, in opioid studies, we often observe count data, where a log-link function may be more appropriate to model the conditional expectation function. We can continue to invoke Assumptions (\ref{asmpt-consistency})-(\ref{asmpt-positivity}) above, but in place of assumption (\ref{asmpt-uniform}), we may instead assume constant multiplicative treatment effects:

\begin{assumption}[Constant multiplicative treatment effects] \label{assume:ConstantMultiplicative}
Letting $\eta_t(\bar{a}; \theta_t)$ be a constant treatment effect for treatment sequence $\bar{a}$ that is a function of parameters $\theta_t$ such that $\eta_t(\bar{0}; \theta_t) = 0$, we have
\( \displaystyle 
\frac{Y_{it}(\bar{a})}{Y_{it}(\bar{0})} = \exp(\eta_t(\bar{a}; \theta_t)).
\)
\end{assumption}

\noindent Analogous to the L-DAM, we can then parameterize the model using $\eta_t(\bar{a}; \theta_t) = \theta^\top\bar{a}$:
\begin{align*}
\log(\mathbb{E}[Y_{it} \mid \bar{Y}_{i,t-1}, \bar{A}_{it}]) &= \alpha + \sum_{b=1}^k \left\{\beta_b \left(\log(Y_{i,t-b}) - \sum_{z=0}^l \theta_{z}A_{i,t-1-z} \right) \right\}
+ \sum_{z=0}^l \theta_{z}A_{i,t-z}.
\end{align*}
\noindent Under constant multiplicative treatment effects, we have that 
$$Y_{it} = Y_{it}(\bar{0}) \exp(\eta_t(\bar{A}_{it}; \theta_t)) = Y_{it}(\bar{0}) \exp(\theta^\top\bar{A}_{it}),$$ which implies that $\log(Y_{it}(\bar{0})) = \log(Y_{it}) - \theta^\top\bar{A}_{it}$. This reveals that in the autoregressive model, we are controlling for the log of counterfactual outcomes, given by $Y_{i,t-b}(\bar{0})$. Additionally, the estimand of interest may change to one on the multiplicative scale, depending on the application of interest, leading to
\begin{align*}
\tilde{\theta}_t(\bar{a}) = \frac{\mathbb{E}[Y_{it}(\bar{a})]}{\mathbb{E}[Y_{it}(\bar{0})]}.
\end{align*}
\noindent The log-linear model and multiplicative estimand add no additional complications. We may continue to use our model to obtain estimates of $\mathbb{E}[Y_{it}(\bar{a})]$ and $\mathbb{E}[Y_{it}(\bar{0})]$ by marginalizing the model over the empirical covariate distribution, setting $\bar{A} = \bar{a}, \bar{0}$, and using the averages of these predictions to compute the estimates of causal contrasts. One potential issue arises when there are zero counts in the data (as $\log(0)$ is undefined). In these settings, we can adjust for a different function of the lagged outcomes other than the log, possibly using basis function expansions to let the data inform the shape of this function.

\section{Connections to existing identification strategies}
\label{sec:Connections}

We briefly compare our identification strategy to other strategies commonly used in causal inference with panel data. First, we compare our approach to difference-in-differences and synthetic controls, commonly used in applied policy applications. Second, we compare it to the sequential g-formula, which identifies causal quantities under sequentially randomized trials and is a more commonly used approach in epidemiological studies.

\subsection{Difference-in-differences and synthetic controls}\label{ssec:did}

Both difference-in-differences and synthetic control analyses are frequently motivated by assuming that the potential outcomes absent treatment follow a linear factor model,

\begin{align*}
Y_{it}(0) = \blambda^\top\bmu + \epsilon_{it},
\end{align*}

\noindent where $\blambda$ is an $r$ by $1$ vector of so-called ``factors'' and $\bmu$ is an $r$ by $1$ vector of so-called ``factor loadings''. Alternatively, one can think of $\blambda$ as a time-varying coefficient vector applied to the unobserved covariate vector $\bmu$. Selection into treatment is then assumed also to be a function of the unobserved covariates $\bmu$. The two-way fixed effects model that often motivates difference-in-differences is a special case of this model where $r = 2$, $\blambda = [1, \lambda_t]$ and $\bmu = [\mu_i, 1]$. One may also extend this model to include observed covariates $X_{it}$ as done in \citep{abadie2010synthetic}.

While we do not observe the unobserved confounder vector $\bmu$, we do observe the pre-treatment outcomes, which are a function of these confounders. Thus, the pre-treatment outcomes may serve as proxy variables for these unobserved confounders. Motivated by this insight, the synthetic controls method estimates the post-treatment counterfactual outcomes absent treatment for a treated unit by reweighting a set of control units to match the mean outcomes of the treated units in the pre-treatment period, and then using the weighted post-treatment outcomes of the control units for the estimate of the potential outcomes under control in the post-treatment period for the treatment group. \cite{abadie2010synthetic} show that under some conditions, as the number of pre-treatment time periods $T_0$ increases, the bias of the synthetic controls estimator approaches zero. Intuitively, the pre-treatment outcomes act as a proxy for the unobserved variables, and the more proxies we observe and adjust for, the better we adjust for the unobserved confounding.

We may similarly motivate our proposed method using the linear factor model, again taking the view that the potential outcomes absent treatment serve as proxies for unobserved confounders. However, in contrast to synthetic controls, our method uses both the pre- and post-treatment data to control for these variables. That is, we use both any observed pre-treatment outcomes and post-treatment estimates of the counterfactual outcomes absent treatment as proxies for these same variables. Assuming our models are well-specified, heuristically the bias of our estimator decreases with the number of lagged potential outcomes absent treatment -- a function of $T$ -- rather than the number of pre-treatment outcomes $T_0$. We prove this result in Proposition \ref{prop:2} in Appendix \ref{app:bound}, albeit under the assumption that the potential outcomes absent treatment are observed variables. One may therefore view our approach as a method to estimate causal effects under the linear factor model by imposing structure on the treatment effect to obtain post-treatment counterfactual estimates absent treatment. 

Our proposed approach also has other benefits. In contrast to synthetic controls, our method more naturally accommodates multiple treated units and staggered treatment adoption. Additionally, in contrast to both synthetic controls and difference-in-differences, our method imputes both the potential outcomes absent treatment for the treated units -- and the potential outcomes under treatment for the control units. This follows because our method imposes structure on the treatment effect that allows us to impute both counterfactuals. As a result, our method targets the average treatment effect (ATE), not simply the average treatment effect on the treated (ATT).

While the linear factor model is one way to motivate the synthetic controls approach, an alternative way is an ignorability assumption that conditions on pre-treatment outcomes \citep{ben2021augmented, abadie2010synthetic}. This implies that treatment is randomized as a function of the pre-treatment outcomes at the time immediately before the first treatment period. Our ignorability assumption again simply generalizes this: we allow the potential outcomes absent treatment to enter the treatment assignment model, but make no restriction on whether these outcomes were actually observed -- that is, whether they occurred pre- or post-treatment. Moreover, this also allows randomization to occur at every time period. In practice, this means that treatment assignment decisions are made with regard to what would have happened in the previous time periods had the intervention not taken place, whether or not the intervention was actually implemented.

\subsection{The sequential g-formula}\label{ssec:gfm}

The sequential g-formula is a more common approach to estimating longitudinal treatment effects in the epidemiological literature, and relies on the so-called ``sequential ignorability'' assumption \citep{gill2001causal}. This assumption states that in every time period, treatment is randomly assigned conditional on all prior outcomes, treatments, and covariates. \footnote{Sequential ignorability states that for all time-periods $t$, $Y_{it}(\bar{a}_t) \perp A_{it} \mid \bar{Y}_{i,t-1}, \bar{X}_{i,t-1}, \bar{A}_{i,t-1}$.} In combination with the corresponding positivity assumptions, sequential ignorability applied to our setting yields the identifying expression (the sequential g-formula), 
{\small
\begin{align}\label{eqn:sgfm}
\mathbb{E}[Y_{it}(\bar{a})] &= \int_{Y_0}...\int_{Y_{T-1}}\mathbb{E}[Y_T \mid \bar{Y}_{T-1}, \bar{A}_{T-1} = \bar{a}_{T-1}] \prod_{t=0}^{T-1}dP(Y_t \mid \bar{Y}_{t-1}, \bar{A}_{t-1} = \bar{a}_{t-1}).
\end{align}
}
\noindent This expression is far more complex than our identification result, which expresses our identified causal parameter in terms of a single averaged conditional expectation function.\footnote{In practice, to estimate equation (\ref{eqn:sgfm}) structure is often placed on $\mathbb{E}[Y(\bar{a}_t)]$; such models are known as marginal structural models \citep{robins2000marginal}.} We obtain a simpler expression because our ignorability assumption conditions on the counterfactual outcomes absent treatment rather than the observed outcomes. To see this, consider a simple case with three time-periods, illustrated in Figure~\ref{fig1} below.
\begin{figure}[H]
\begin{center}
\begin{tikzpicture}
[
array/.style={rectangle split, 
	rectangle split parts = 3, 
	rectangle split horizontal, 
    minimum height = 2em
    }
]
 \node (1) {$Y_0$};
 \node [right =of 1] (2) {$A_1$};
 \node [right =of 2] (3) {$Y_1$};
 \node [right =of 3] (4) {$A_2$};
 \node [right =of 4] (5) {$Y_2$};
 \node [below =of 3] (6) {U};
 \draw[Arrow] (1.east) -- (2.west);
 \draw[Arrow] (2.east) -- (3.west);
 \draw[Arrow] (3.east) -- (4.west);
 \draw[Arrow] (4.east) -- (5.west);
 \draw[Arrow] (6) to (1);
 \draw[Arrow] (6) to (3);
 \draw[Arrow] (6) to (5);
 \draw[Arrow] (1) to [out = 25, in = 160] (5);
 \draw[Arrow] (1) to [out = 25, in = 160] (4);
 \draw[Arrow] (1) to [out = 25, in = 160] (3);
 \draw[Arrow] (2) to [out = 25, in = 160] (5);
 \draw[Arrow] (2) to [out = 25, in = 160] (4);
 \draw[Arrow] (3) to [out = 25, in = 160] (5);
 \color{black}
\end{tikzpicture}\caption{Conventional longitudinal causal structure} \label{fig1}
\end{center}
\end{figure}
\noindent Ideally, we would hope to identify $\mathbb{E}[Y_2(\bar{a})]$ in terms of an averaged conditional expectation function. One possibility is the averaged autoregressive model,
\begin{align*}
\mathbb{E}[\mathbb{E}[Y_2 \mid Y_1, Y_0, A_1 = a_1, A_2 = a_2]].
\end{align*}
\noindent However, equivalence between this expression and $\mathbb{E}[Y_2(\bar{a})]$ would require that $(A_1, A_2) \perp Y_2(a_1a_2) \mid (Y_0, Y_1)$. Unfortunately, this will not hold under the causal structure depicted in Figure \ref{fig1}. The problem, as mentioned previously, is that $Y_1$ acts as a collider between $A_1$ and $Y_2(a_1a_2)$, since $Y_1$ lies on the causal pathway from $A_1$ to $Y_2$. Conditioning on $Y_1$ therefore introduces a dependence between $Y_2(a_1a_2)$ and $A_1$, violating the required ignorability assumption. 

By contrast, our ignorability assumption conditions on $(Y_1(00), Y_0(00))$, variables that do not lie on the causal pathway from $A_1$ to $Y_2$. Figure \ref{fig2} depicts a causal diagram illustrating the assumed relationships between variables, where blue arrows indicate deterministic relationships (notice that the potential outcomes $Y_t(a_1a_2)$, omitted from this diagram, are a deterministic function of $Y_t(00)$ by assumption). Figure~\ref{fig2} implies the conditional independence statement $(A_1, A_2) \perp Y_2(a_1a_2) \mid Y_1(00), Y_0(00)$. By conditioning on the unobserved potential outcomes absent treatment rather than the observed outcomes, our approach avoids collider bias, allowing us to achieve causal identification in terms of an averaged conditional expectation function.

\begin{figure}[H]
\begin{center}
\begin{tikzpicture}
[
array/.style={rectangle split, 
	rectangle split parts = 3, 
	rectangle split horizontal, 
    minimum height = 2em
    }
]
 \node (1) {$Y_0$};
 \node [right =of 1] (2) {$A_1$};
 \node [right =of 2] (3) {$Y_1$};
 \node [right =of 3] (4) {$A_2$};
 \node [right =of 4] (5) {$Y_2$};
\node [below =of 1] (6) {$Y_0(00)$};
\node [below =of 3] (7) {$Y_1(00)$};
\node [below =of 5] (8) {$Y_2(00)$};
\node [below =of 7] (9) {$U$};
 \draw[Arrow] (6.east) -- (7.west);
 \draw[Arrow] (7.east) -- (8.west); 
 \color{blue}
 \draw[Arrow] (2.east) -- (3.west);
 \draw[Arrow] (4.east) -- (5.west);
 \draw[Arrow] (6) to (1);
 \draw[Arrow] (7) to (3);
 \draw[Arrow] (8) to (5); 
 \color{black}
 \draw[Arrow] (6) to (2);
 \draw[Arrow] (7) to (4);
 \draw[Arrow] (2) to [out = 25, in = 160] (4);
 \draw[Arrow] (2) to [out = 25, in = 160] (5);
 \draw[Arrow] (6) to [out = -25, in = -160] (8);
 \draw[Arrow] (9) to (6);
 \draw[Arrow] (9) to (7);
 \draw[Arrow] (9) to (8);
\color{black}
\end{tikzpicture}\caption{Assumed longitudinal causal structure} \label{fig2}
\end{center}
\end{figure}

\noindent However, this identification strategy comes with a cost: because we do not necessarily observe these counterfactual quantities required for ignorability, we instead rely on structural assumptions that parameterize the treatment effects as a function that must be estimated. By contrast, equation (\ref{eqn:sgfm}) makes no such restriction. As a result, our approach allows for simpler estimation methods, but at the cost of stronger modeling assumptions.  

\subsection{Model validation}
Once a model has been specified, when sufficient pre-period data are available, we recommend conducting placebo tests similar to those used in the synthetic controls literature to choose or validate the specification (\cite{abadie2015comparative}). Specifically, assume that we observe $t = 1, ... T_0$ time periods where no state is treated until time $T_0 + 1$. This leaves $T_0$ pre-treatment periods and $T_P = T - T_0$ post-treatment time periods. When $T_0 \ge T_P$, we may take the observed distribution of treatment assignment in the post-treatment period and shift them backwards in time $T_P$ time-periods to generate pseudo-treatments in the pre-treatment period. When $T_0 < T_P$, we could instead move either the first or final $T_0$ post-treatment time-periods back to the pre-treatment period. Finally, we may then run the model using this pre-period data and pseudo-treatment time periods to estimate the pseudo-treatment effects to test the models. 

Specifically, when the pre-period estimates of $\theta$ are statistically indistinguishable from zero, this provides additional support to a given model. By contrast, a non-null effect may suggest a poorly specified model that may be revised, for example, by adding additional lagged outcomes or covariates to the specification.\footnote{Notice, however, that using pre-treatment data to conduct model selection rather than model validation may result in overfitting. To conduct model selection, one may divide the pre-treatment data into two sets to first select then to validate the model \citep{abadie2015comparative}.} The value of these placebo tests in this context, however, is slightly more limited than for synthetic controls or difference-in-differences applications. The goal of synthetic controls and DiD is to predict $Y_{it}(0)$ for the post-treatment treated units. However, our model must predict both $Y_{it}(0)$ for the post-treatment treated units \textbf{and} $Y_{it}(1)$ for the post-treatment control units. Heuristically, because we observe $Y_{it}(0)$ for the pre-treatment treated units, the pre-treatment data is arguably informative about the model of $Y_{it}(0)$ during the post-treatment period. However, this same logic does not hold for the post-treatment control units, whose values of $Y_{it}(1)$ are unobserved in the pre-treatment period. That is, because we do not observe $Y_{it}(1)$ for any units in the pre-treatment period, we also cannot model $Y_{it}(1)$ for any units, so that this placebo test does not inform any part of the model corresponding to the treatment effect. Despite this potential limitation, we believe these are still useful tests to implement and help users ensure their chosen statistical model is well-suited to their application. 

\section{Simulation studies}
\label{sec:Simulations}

We evaluate the performance of the proposed approach against state-of-the-art methods for panel data causal inference in a simulation study. We explore four different estimators in this manuscript: 1) The linear autoregressive model described in Sections \ref{sec:methods} and \ref{sec:relaxations}, 2) The two way fixed effects estimator, 3) the synthetic control estimator for staggered adoption developed in \cite{ben2022synthetic}, and 4) the difference in differences estimator for staggered adoption developed by \cite{callaway2021difference}. In Section \ref{ssec:SimOpioid}, we compare these approaches on simulations using the observed opioid data. Then, in \ref{ssec:SimArtificial} we use fully simulated data generated under linear factor models to evaluate the performance of our estimator in the presence of time-varying unmeasured confounding. 

\subsection{Results on simulation using opioid prescribing data}
\label{ssec:SimOpioid}

Instead of simulating outcomes from a data generating process we fully specify, we first utilize existing outcome data from our application to ensure that the results reflect performance in realistic data and are more relevant to our setting. Specifically, we let our observed outcome data represent $Y_{it}(0)$, and then we simulate treatment effects to generate $Y_{it}(1)$. Our data consist of 51 units and 16 time periods, and we choose $n_t = 20$ units to be treated in all simulations. Our observed outcomes are the number of opioids prescribed by each state (and Washington DC) and year. We do not vary the outcome values across simulations, but instead vary which units are treated, when they are treated, and the heterogeneity of the causal effect (when applicable). Specific details about the simulation scenarios can be found in Appendix D, though we summarize the key points here. We let treatment status at time $t$ depend on $Y_{i,t-1}(0)$, and simulate a staggered adoption setting where a unit remains treated once they are treated. We vary the degree of confounding by $Y_{i,t-1}(0)$ into three categories we refer to as no confounding, moderate confounding, and high confounding. We set a small average treatment effect and vary the degree of heterogeneity of this effect across units, split into categories denoted by no heterogeneity, moderate heterogeneity, and high heterogeneity. The moderate and high heterogeneity settings represent scenarios where the probability that a unit has an individual treatment effect either twice as large, or the opposite sign, as the average treatment effect is set to be 0.1 and 0.33, respectively. It is also important to emphasize that we do not simulate the outcomes and therefore they may or may not fit well into the autoregressive framework we presented. Throughout this section, we adjust for the prior two time periods' counterfactual outcomes. We also use the linear autoregressive model here that uses a normal approximation to the rates of the outcome instead of directly modeling counts as described in Section \ref{ssec:Nonlinear}. In Appendix D we run the same simulations using negative binomial autoregressive models and find largely similar results. 

To ensure a fair comparison across different estimation procedures, our target of inference throughout is the sample treatment effect on the treated described in Section \ref{sec:methods} -- this estimand is easily estimated using any of the synthetic control, DiD, and autoregressive models we consider. Interpretation of the two-way fixed effects (TWFE) estimator may be complicated in this setting \citep{imai2021use}, but we include it due to its widespread use. We calculate relative mean squared error (MSE), which in any simulation scenario is the MSE of an estimator divided by the minimum MSE across all estimators, leading to the optimal estimator having an MSE of 1. We also calculate 95\% interval coverage rates and the power to detect a nonzero effect, which is calculated as the proportion of 95\% intervals for which zero is not included and the interval correctly estimates the sign of the treatment effect. 

\begin{figure}[htbp]
    \centering
    \includegraphics[width=0.8\linewidth]{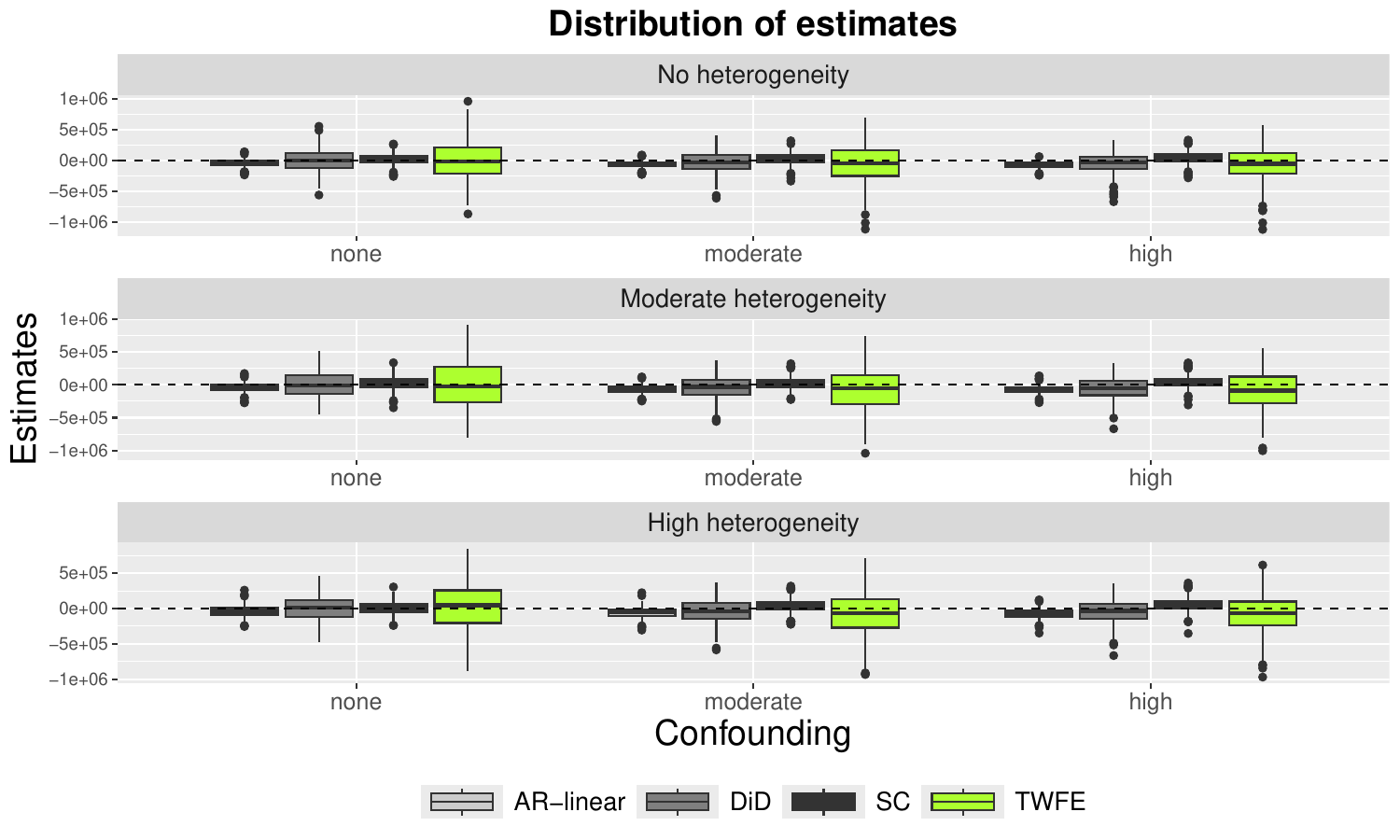}
    \caption{Distribution of point estimates for simulations on opioid prescribing data.}
    \label{fig:33results}
\end{figure}

\begin{figure}[htbp]
    \centering
    \includegraphics[width=0.8\linewidth]{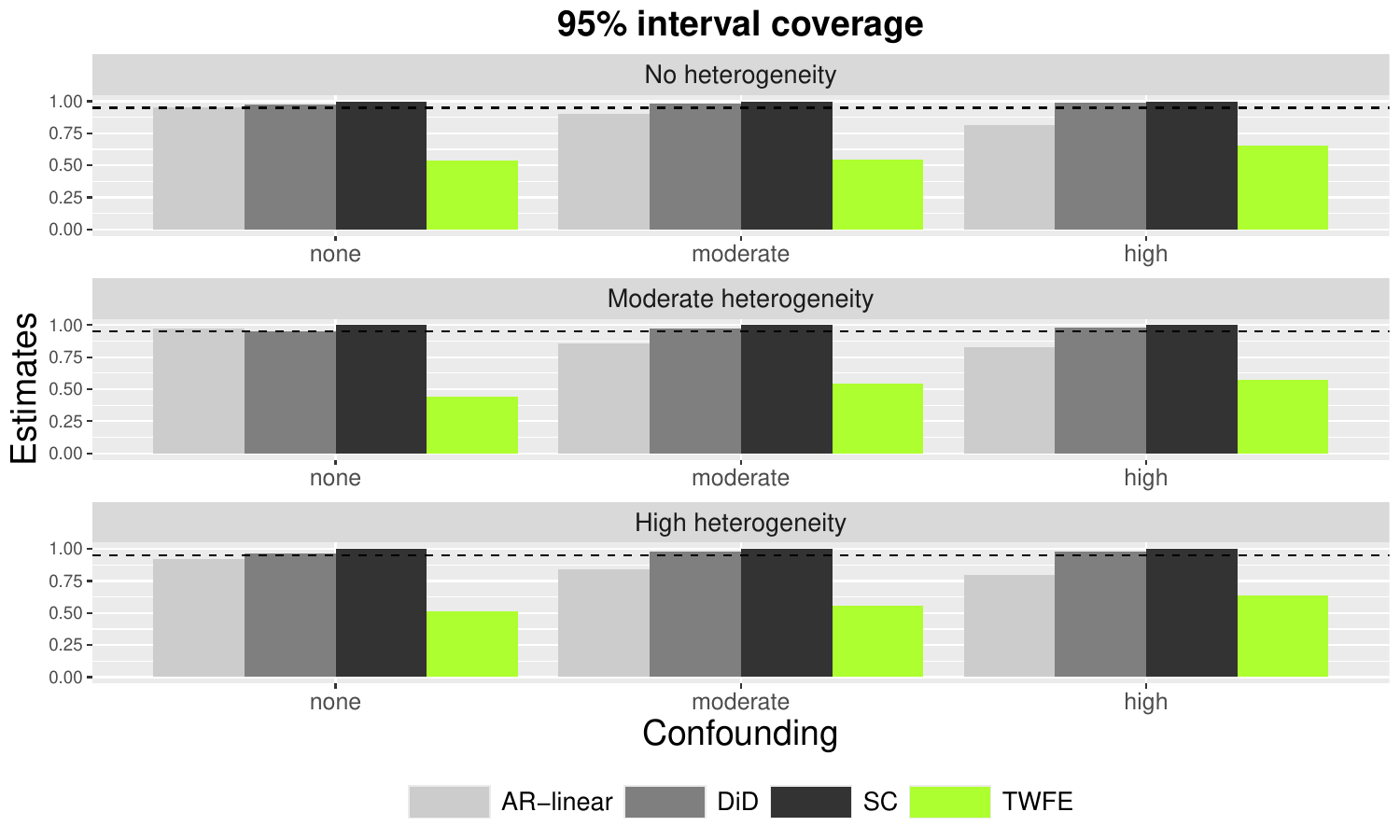}
    \caption{95\% interval coverage for simulations on opioid prescribing data.}
    \label{fig:33resultsCoverage}
\end{figure}

\begin{figure}[htbp]
    \centering
    \includegraphics[width=0.8\linewidth]{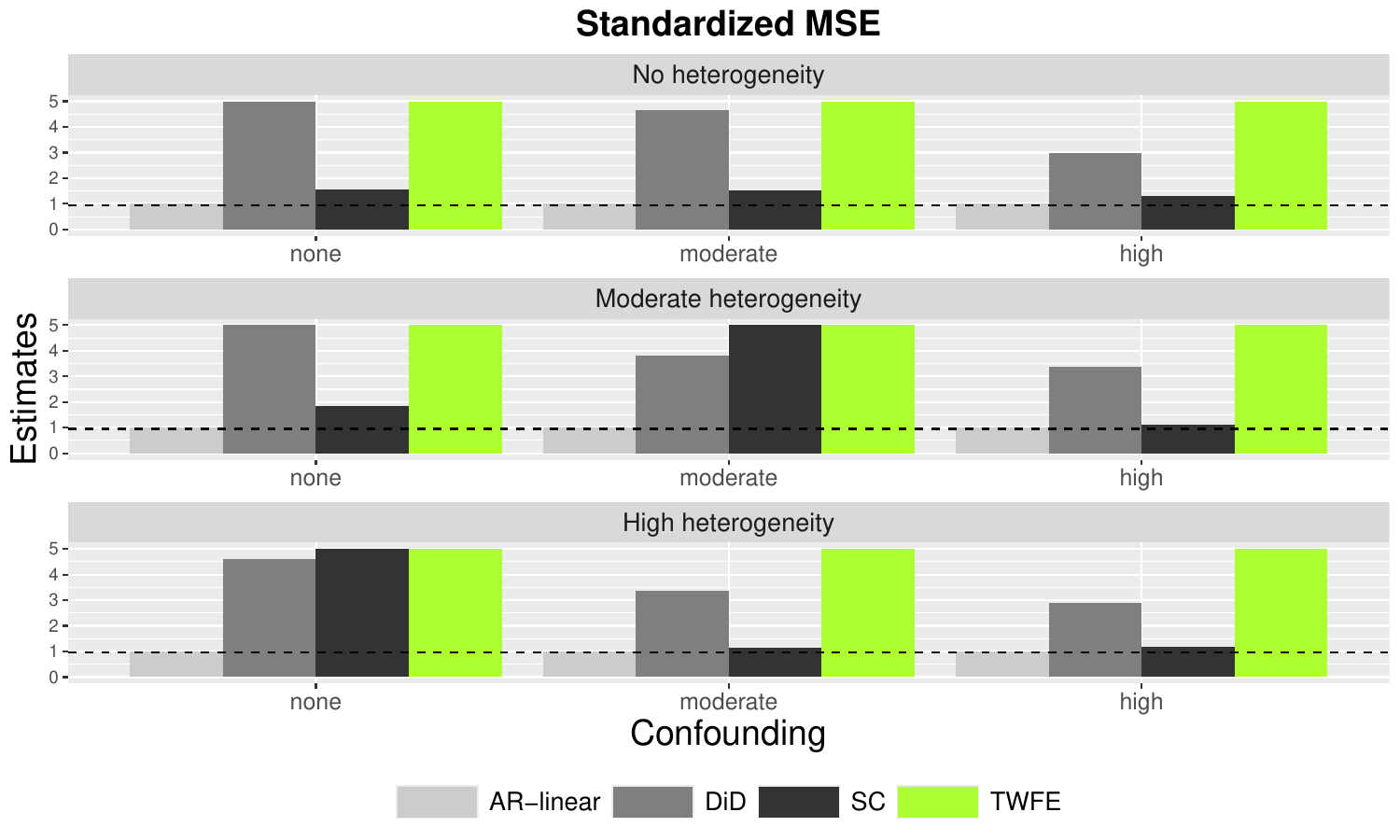}
    \caption{Mean squared error for simulations on opioid prescribing data. MSE values have been capped at 5 for easier comparison.}
    \label{fig:33resultsMSE}
\end{figure}

\begin{figure}[htbp]
    \centering
    \includegraphics[width=0.8\linewidth]{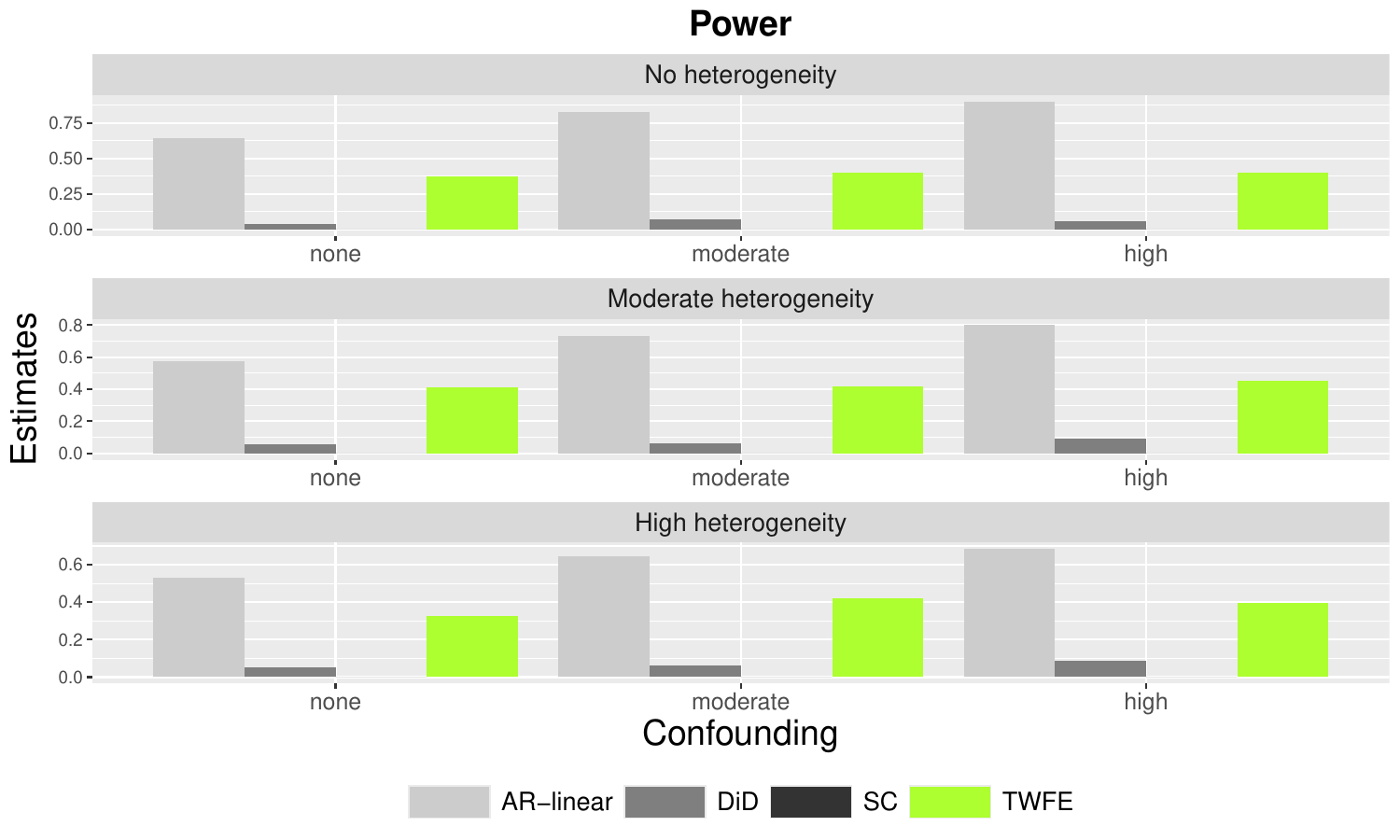}
    \caption{Power for simulations on opioid prescribing data.}
    \label{fig:33resultsPower}
\end{figure}

Figures \ref{fig:33results} - \ref{fig:33resultsPower} display the results. The best performing estimators are the autoregressive model and synthetic control approach, followed by the difference in differences estimator, and then the TWFE estimator. The 95\% interval coverage is good for the autoregressive, DiD, and SC approaches with the autoregressive approach only falling below 95\% coverage with high heterogeneity or large amounts of confounding. The autoregressive approach has the best MSE in all scenarios with a relative MSE of 1, and it is substantially more efficient in certain scenarios. This highlights the bias-variance trade-off inherent to using these models as they may be significantly more efficient than existing estimators that are not model based, although they may be more sensitive to model misspecification. This boost in efficiency is most notable in Figure \ref{fig:33resultsPower}, which displays power. Both the DiD and SC approaches have effectively no power to detect the nonzero treatment effects, while the autoregressive approach detects this effect in over half of the simulated data sets. This is crucial, because policy effects are typically very small in state-level opioid policies and it appears that existing approaches are not well-suited to detecting these effects. Another key takeaway from the simulation is that reasonable amounts of heterogeneity do not appear to negatively impact the performance of the autoregressive model, consistent with the results of Section \ref{sec:Heterogeneity}. 

\subsection{Results using fully simulated data}\label{ssec:SimArtificial}

\begin{figure}[htbp]
    \centering
    \includegraphics[width=0.8\linewidth]{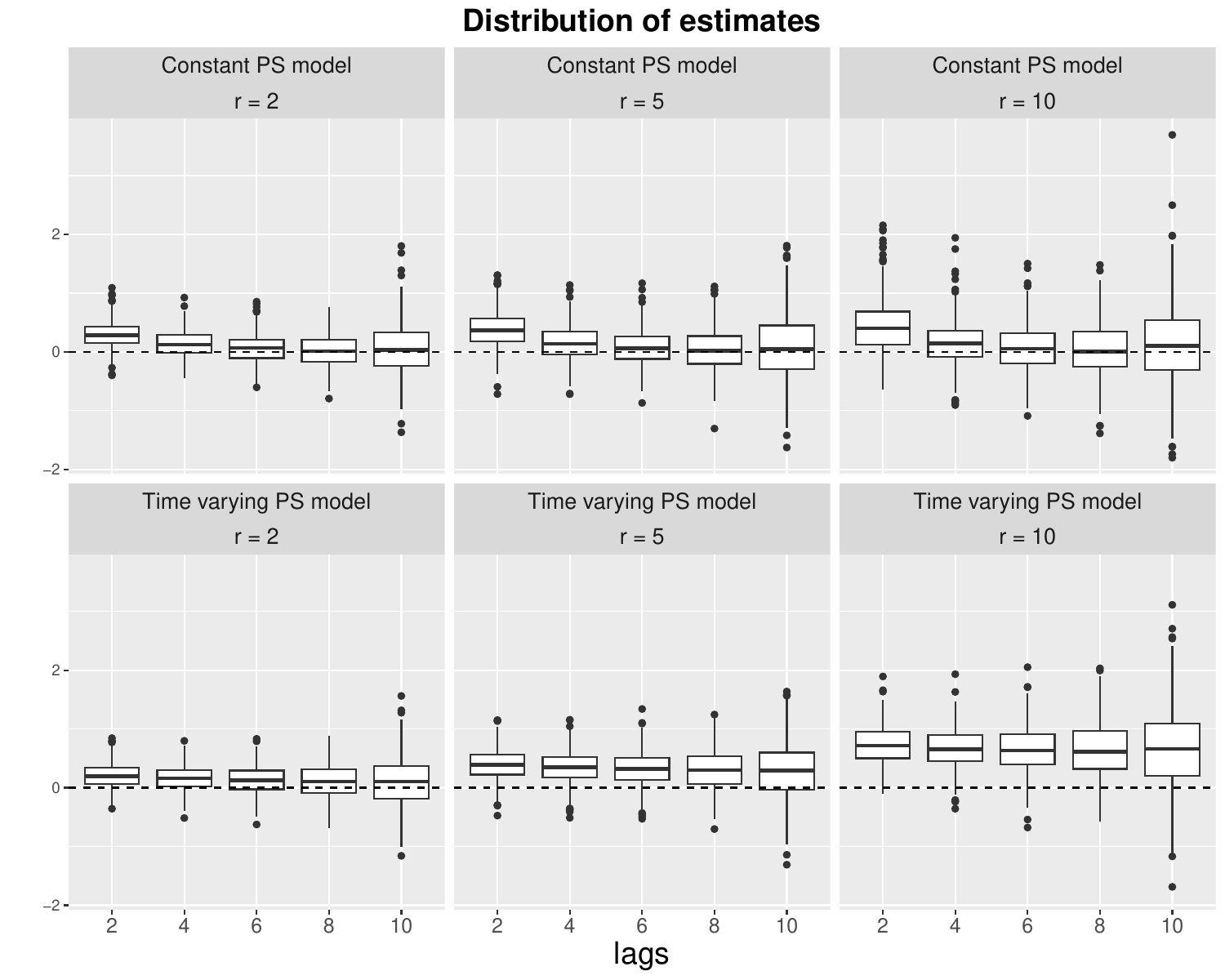}
    \caption{Distribution of point estimates for simulations from latent factor models. }
    \label{fig:FactorSims}
\end{figure}

We also present results using fully simulated data to highlight the performance of our approach under the linear factor model that was described in Section \ref{ssec:did}, and that is commonly used to justify synthetic control approaches, and that nests difference-in-differences as a special case. Specifically, we generate data from a factor model given by
\begin{align*}
Y_{it}(0) = \blambda\bmu + \epsilon_{it},
\end{align*}
\noindent where we generate the latent factors $\blambda$ from independent standard normal distributions. The time-varying coefficients $\blambda$ are drawn randomly from a standard uniform distribution, and $\epsilon_{it} \stackrel{iid}\sim \mathcal{N}(0,1)$. The number of latent factors is given by $r$, which we vary in $\{2, 5, 10\}$ to show the impact of increasing numbers of latent factors. Throughout, we set the number of units to be $n=20$, which are measured over $T = 12$ time points. We explore two distinct simulation scenarios using the factor model above, which vary in the treatment assignment mechanism. Note that treatment status for one unit in this scenario can switch between the treatment and control condition over time unlike in the previous simulation setting. Letting $g(\cdot)$ be the standard logistic link function, these two propensity score models are given below.
\begin{enumerate}
    \item A constant assignment model with probabilities given by
    $$P(A_{it} = 1 \mid \bmu) = g \bigg(-0.5 + \sum_{j}\mu_{ij} \bigg)$$
    \item A time-varying assignment model where $\boldsymbol{\xi}_t$ are randomly assigned from independent, standard uniform distributions, and the propensity score is given by
    $$P(A_{it} = 1 \mid \bmu) = g \bigg(-0.5 + \boldsymbol{\xi}_t^\top \bmu \bigg)$$
\end{enumerate}
We fit our proposed approach and vary the number of lags that we adjust for in $\{2, 4, 6, 8, 10\}$. The point estimates across 500 replications can be found for each scenario in Figure \ref{fig:FactorSims}. 

Our approach is able to obtain approximately unbiased results in the constant propensity score setting, but is unable to achieve these same results in the time-varying propensity score setting. Moreover, the bias in the constant propensity score setting decreases with the number of lags included in the model, although we increasingly lose efficiency, reflected by the increasing spread of the point estimates. The bias results are consistent with our results in Proposition \ref{prop:2} in Appendix \ref{app:bound} and with \citep{abadie2010synthetic}. Heuristically, the failure of our model in the time-varying setting occurs because the distribution of factors across the treatment and control groups change over time; as a result, controlling for the prior counterfactual outcomes cannot simultaneously balance all unobserved factors across the treatment and comparison groups in each time period.\footnote{We expect that this failure would occur in any model that does not either implicitly or explicitly allow the treatment assignment model to vary over time.} By contrast, in the constant propensity score setting, this distribution remains constant over time, and the lagged outcomes appropriately proxy for the unobserved factors. Moreover, the more lags we control for, the better our models balance these factors appropriately across the treatment and comparison groups.

\section{Examining the impact of state policies on opioid prescriptions}\label{sec:Application}

We now estimate the effect of the four state-level policies described in Section \ref{sec:DataDescription} on the number of opioid prescriptions per capita in each state over the years 2006-2017. We first focus on estimating the effects of single policies and compare effect estimates with existing approaches to inference in this problem. We then estimate the effect of multiple policies simultaneously using the proposed autoregressive framework. 

\subsection{Single policy analysis}

First, we analyze the impact of each policy on prescribing separately, which facilitates comparison with existing approaches designed for one policy only. Additionally, we have information about the proportion of each year that the policy is implemented for each state, but for this analysis, we dichotomize policy variables based on whether the state has any law in place that year. This allows for implementation of existing methods designed for binary treatments, but it is worth noting that our approach applies even when $A_{it}$ is a proportion. We explore this further in the multiple policy analysis in Section \ref{ssec:AnalysisMultiplePolicy}. 

\subsubsection{Model formulation and estimands of interest}

We estimate the sample average treatment effect on the treated, which represents the average impact of the policy on the treated units in the first 3 years it is enacted. We utilize four distinct estimators of this quantity: 1) a DiD estimator allowing for staggered adoption, 2) the augmented synthetic control approach that allows for staggered adoption, 3) the autoregressive model incorporating time-varying covariates, and 4) the autoregressive model without time-varying covariates. For the autoregressive models, we must specify the specific form of our model as well as the prior distributions to be used. We assume that $k=2$ and that a second order autoregressive model is sufficient, and we enforce that only the policy indicator at the most recent three time points affects the outcome. The autoregressive coefficients are assigned standard uniform prior distributions, while all regression parameters, including those that govern the treatment effect, are assigned noninformative normal priors centered at zero with a variance of 100, and the variance parameters are assigned inverse gamma priors with parameters 0.001 and 0.001. 

\subsubsection{Estimates of average causal effects}
\label{sssec:AnalysisAverage}
The results for each of the four policies under each of the four estimators is given in Table \ref{tab:SinglePolicyEstimates}. Note there is no estimate for the augmented synthetic control estimator for the medical marijuana policy, as many states had already enacted this policy at the start of our study and therefore synthetic controls cannot estimate this policy effect. For both NAL policies as well as medical marijuana, the results are fairly consistent, showing that they do not have an impact on opioid prescriptions. All estimators estimate a negative effect of PDMP laws on prescriptions suggesting must access PDMP laws decrease opioid prescribing, with both the DiD and autoregressive models showing a statistically significant effect. In general across all four policies, the autoregressive models tend to have much smaller interval widths, highlighting the efficiency gains of this approach. 

\begin{figure}[H]
    \centering
        \includegraphics[width=0.95\textwidth]{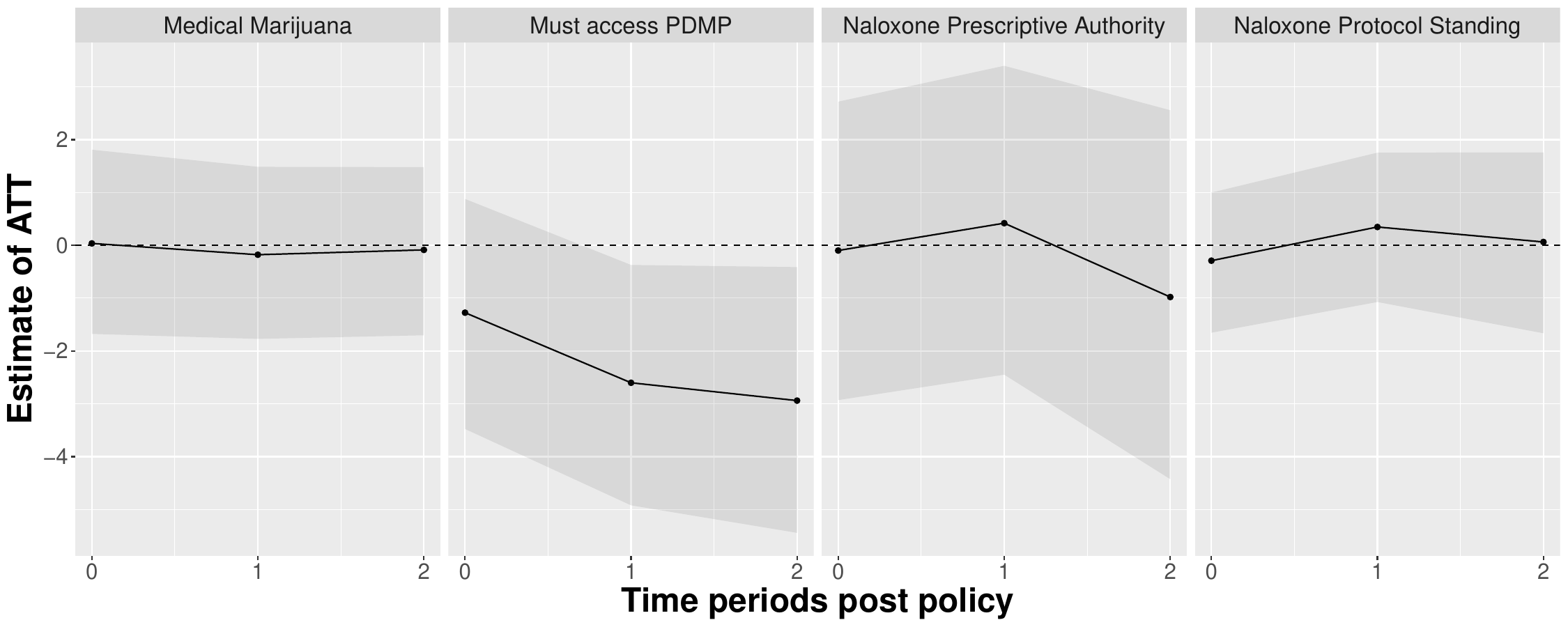}
    \caption{Estimates, and corresponding 95\% credible intervals, for the time-specific sample average treatment effect on the treated from the autoregressive model that includes covariates. }
    \label{fig:SinglePolicyTimeVarying}
\end{figure}

In addition to the sample average treatment effect, we also estimate a time specific sample average treatment effect that does not average over time periods. Figure \ref{fig:SinglePolicyTimeVarying} shows estimates of this quantity using the autoregressive model, revealing no effect at any time period for NAL or medical marijuana laws. The effect of PDMP laws, however, becomes more pronounced after the first year. This may be due to the fact that we dichotomized the treatment variable, but in the first year in which it is enacted, it is commonly only enacted for a fraction of the year. 

\begin{table}[ht]
\centering
\begin{tabular}{rllll}
  \hline
 & DiD & Augmented SC & AR & AR no covariates \\ 
  \hline
NAL protocol standing & -0.50 (-3.06, 2.05) & -0.20 (-6.03, 5.31) & 0.04 (-0.92, 0.94) & 0.02 (-0.89, 0.91) \\ 
  NAL prescriptive authority & 0.32 (-2.45, 3.09) & 1.35 (-5.61, 8.11) & -0.22 (-2.09, 1.6) & -0.25 (-2.21, 1.59) \\ 
  Must Access PDMP & -4.77 (-8.87, -0.68) & -3.86 (-9.77, 1.58) & -2.27 (-3.7, -0.85) & -2.29 (-3.72, -0.84) \\ 
  Medical marijuana & 1.34 (-1.23, 3.91) & Not available & -0.08 (-1.16, 0.99) & -0.13 (-1.12, 0.87) \\ 
   \hline
\end{tabular}
\caption{Estimates and 95\% intervals for the sample average treatment effect on the treated across all estimators. }
\label{tab:SinglePolicyEstimates}
\end{table}

\subsubsection{Exploring treatment effect heterogeneity}

We also explore effect heterogeneity, using the methodology described in Section \ref{sec:Heterogeneity} and let $V_{it} = X_{it}$ so that we study heterogeneity across all observed covariates. First, we examine whether accounting for heterogeneity leads to different average causal effect estimates than those in Section \ref{sssec:AnalysisAverage}. The estimates are largely similar. The average effects and corresponding credible intervals are for each policy are: 1) NAL protocol standing: 0.36 (-0.69, 1.47); 2) NAL prescriptive authority: -0.26 (-2.31, 1.78); 3) Must Access PDMP: -2.79 (-4.72, 1.02); and 4) Medical marijuana: -0.03 (-1.14, 1.10). While these yield the same substantive conclusions as before, the widths of the 95\% credible intervals for all four policies are wider, illustrating the trade-off from the additional model complexity. 

\begin{figure}[htbp]
    \centering
    \includegraphics[width=0.8\linewidth]{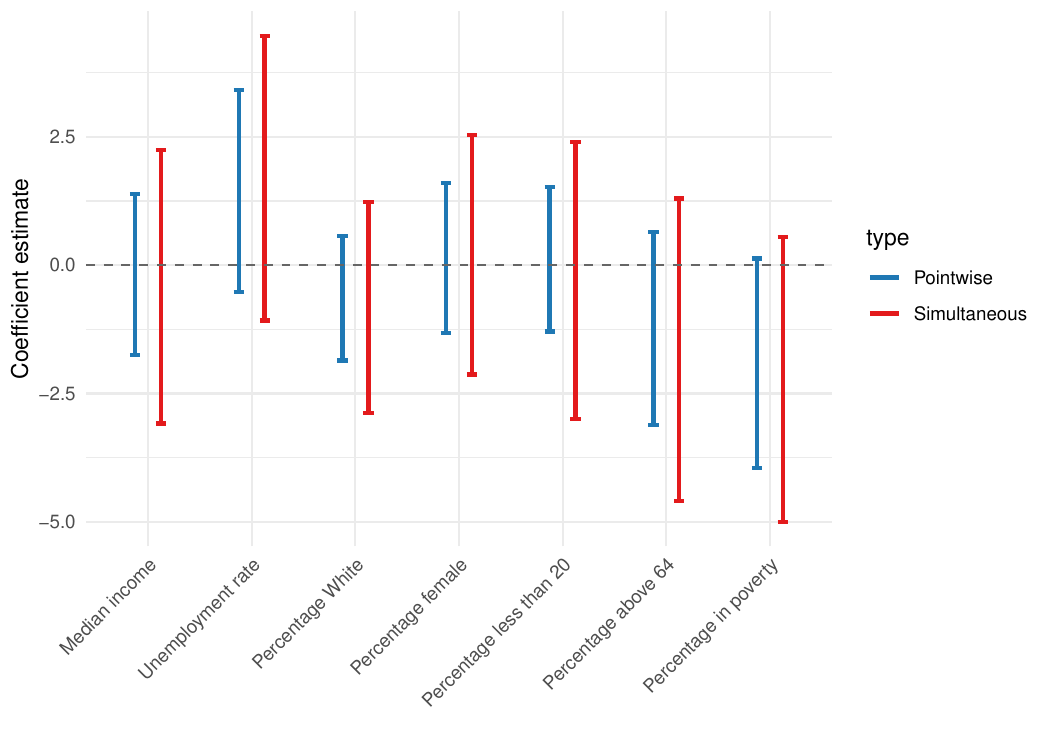}
    \caption{95\% credible intervals for the coefficients of the treatment effect heterogeneity function when examining Must Access PDMP policies. }
    \label{fig:HeterogeneityCoefficients}
\end{figure}

We also examine the extent of heterogeneity for any of the policies by examining the coefficients $\nu_j$, which highlight how the treatment effect function changes with each covariate. We show these coefficients for the Must Access PDMP policy, which is the only policy to show a marginal effect on opioid rates. Figure \ref{fig:HeterogeneityCoefficients} displays the results. All intervals, whether pointwise or simultaneous, contain zero, indicating that we cannot reject the null hypothesis of no treatment effect heterogeneity. The interval for percentage in poverty is mostly negative, indicating that states with higher numbers of people in poverty would have more pronounced, negative effects of must access PDMP policies. 

\subsection{Examining multiple policies simultaneously}
\label{ssec:AnalysisMultiplePolicy}
We may also use the proposed approach to estimate the effect of multiple policies simultaneously, allowing us to see whether certain policies are more (or less) effective if implemented in tandem with other policies. We no longer dichotomize treatment variables and instead use the proportion of that year that the policy was implemented as our main treatment variables. We use a similar model formulation as for the single policy case as we utilize $k=2$ autoregressive lags, though we now only allow the treatments at the current time point to affect outcome. We do this partially to reduce the complexity of the model to allow us to study multiple policies and their interactions simultaneously. For this analysis, we focus on two policies of interest: must access PDMPs and medical marijuana laws. We focus on the must access PDMP and medical marijuana laws specifically, because they are the two most likely to affect opioid prescription rates. Naloxone access laws are primarily expected to reduce opioid mortality by intervening in cases of an overdose, rather than by reducing prescriptions.\footnote{Studying all four laws simultaneously is exceedingly difficult with only 51 units and leads to serious issues with positivity violations as there are combinations of the four policies for which we have no data.}

The estimand of interest is similar to the one studied for single policy analyses, but now potential outcomes are denoted by $Y_{it}(a, m)$, which is the outcome we would observe for unit $i$ at time $t$ had they been exposed to levels of the two policies given by $a$ and $m$. We look at a sample average potential outcome given by $SAPO(a, m) = \frac{1}{n} \sum_{i=1}^n  Y_{it}(a, m)$, for all combinations of $a$ and $m$. We focus on $t=6$, which corresponds to the year 2011, though similar results are obtained examining different years. This shows the average outcome, across all states, for each combination of the two policies. Moving forward, the first policy will represent must access PDMP laws, while the second policy will represent medical marijuana laws. 

\subsubsection{Results}
\begin{figure}[H]
    \centering
        \includegraphics[width=0.8\textwidth]{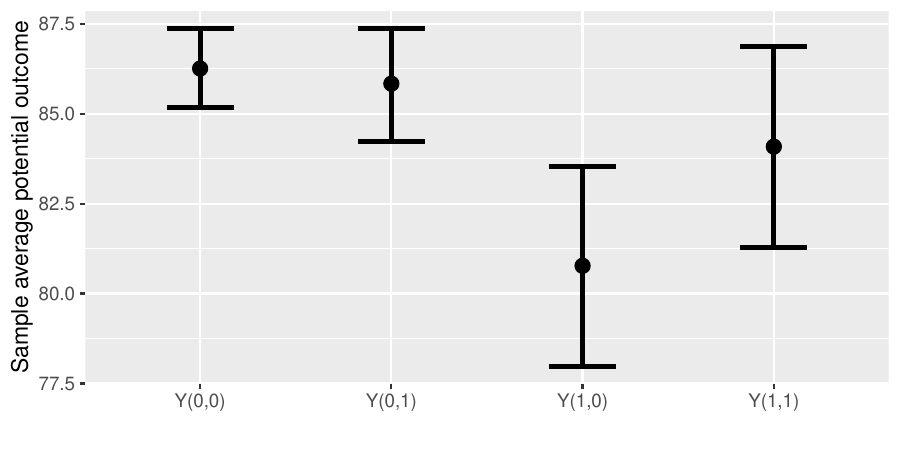}
    \caption{Estimates of SAPO for opioid prescriptions under all combinations of PDMP and medical marijuana laws}
    \label{fig:MultiplePolicySATE}
\end{figure}

The estimates and corresponding 95\% credible intervals for SAPO($a$, $m$) for the four combinations of policies is given in Figure \ref{fig:MultiplePolicySATE}. The values of SAPO($0$, $0$) and SAPO($0$, $1$) are both high, indicating that medical marijuana does little to reduce opioid prescriptions. SAPO($1$, $0$) is much lower, which re-iterates what was seen in the single policy analysis, which is that must access PDMP laws alone are able to substantially reduce opioid prescriptions at the state level. Interestingly, the value of SAPO($1$, $1$) is moderately higher than SAPO($1$, $0$), suggesting that the presence of medical marijuana policies can dampen the influence of must access PDMPs. To see this, we may examine the posterior distribution of the coefficients corresponding to the policy variables in the autoregressive model. The treatment effect component of the model is given by $A \theta_{pdmp} + M \theta_{mm} + (A \times M) \theta_{int}$, and the posterior distribution of these three parameters can be seen in Appendix E. The interaction term has posterior density almost entirely greater than 0 (95\% CI: 0.50, 4.89), highlighting that medical marijuana laws temper the effect of the must access PDMP laws on opioid prescriptions. 

\section{Discussion}\label{sec:Discussion}

We examined the impact of four policies on opioid prescribing, highlighting meaningful evidence that must access PDMPs decreases opioid-prescribing rates within a state. We also found null effects showing that naloxone access laws and medical marijuana did not meaningfully reduce opioid prescriptions at the state level, and that medical marijuana may reduce the impact of must access PDMPs when enacted jointly. Taken together, our findings suggest that a combination of policies may be needed to effectively address the opioid crisis. In the end, the application must be interpreted with caution when considering the null findings for the other policies. The policies we examine were enacted to improve the lives of individuals in the United States, and even small effects of policies could translate to saving or helping thousands of individuals. 

Our work also makes a strong methodological contribution. In particular, we propose a novel causal framework that provides formal justification for using debiased autoregressive models to estimate causal effects. While our causal assumptions make strong restrictions on effect heterogeneity, we show mathematically that the presence of effect heterogeneity likely only causes a small bias. Moreover, we show how our approach can allow for identification and estimation in the presence of multiple policy changes. By contrast, effect estimation in this context is more challenging for difference-in-differences and synthetic controls methods. Finally, we provide simulation evidence showing that our proposed method outperforms existing methods in opioid-policy settings where we have small sample sizes with a changing policy landscape, consistent with previous studies \citep{griffin2021moving,schell2018evaluating,griffin2023confounding,griffin2022cooccur}. More generally, the proposed debiased autoregressive models offer a promising method for policy researchers to evaluate the effects of social and health policies in multiple applied areas, including gun violence, opioids, and COVID-19 policy evaluations. By providing relatively low variance effect estimates in small data settings, this method allows for more precise causal inferences with relatively low bias.

On the other hand, this method is not without limitations.  Our identification and estimation strategy requires making strong functional form assumptions on the treatment effect function itself, a problem that may be difficult to reason about in certain applications. Additionally, while we have shown that effect heterogeneity is likely only to induce small bias with respect to effect estimation in a highly simplified setting, we have not formally shown that these results extend in more complex settings where we condition on multiple pre-treatment outcomes, covariates, and where we attempt to evaluate the effects of multiple policies. Valuable future research may therefore consider how to conduct formal sensitivity analyses with respect to the required causal assumptions, and to formalize how heterogeneity may affect model-based estimates in more complex settings. Nevertheless, our results make strong progress in formally justifying the use of debiased autoregressive models for causal effect estimation and for their utility in applied policy settings. 

\section*{Acknowledgements}
This research was financially supported through a National Institutes of Health (NIH) grant (P50DA046351) to RAND (PI: B.D.S.) as well as Arnold Ventures (AV). NIH and AV had no role in the design of the study, analysis, and interpretation of data nor in writing the manuscript. We wish to thank Andrew Morral, Megan Schuler, Bradley Stein, Rosalie Pacula, David Powell, Mark Sorbero, and Evan Peet for comments on different iterations of this paper and/or the execution of the policy evaluation. We also thank the members of the OPTIC Advisory Board for their feedback on these analyses.

\bibliographystyle{imsart-nameyear} 
\bibliography{biblio}       

\newpage

\appendix

\section{Causal assumptions for multiple policies}

We define our target parameter as the counterfactual quantity $\mathbb{E}[Y_{it}(\bar{a}, \bar{m})] - \mathbb{E}[Y_{it}(\bar{a}', \bar{m}')]$. To identify this quantity in terms of the observed data, we first assume consistency,

\begin{assumption}[Consistency]\label{asmpt-1a}
\begin{align*}
\bar{A}_{it} = \bar{a}, \bar{M}_{it} = \bar{m} \implies Y_{it} = Y_{it}(\bar{a}, \bar{m}), \qquad \forall \bar{a}, \bar{m}.
\end{align*}
\end{assumption}

\noindent This assumption extends the previous assumption to account for multiple policies. Similar to before, this assumption precludes interference across states (or units) -- that is, each states' observed outcome is only a function of the policies implemented in that state. We next assume $l$-limited carryover effects.

\begin{assumption}[$l$-limited carryover effects]\label{asmpt-2a}
Let $\bar{a}, \bar{m}$ and $\bar{a}', \bar{m}'$ be two treatment sequences such that $a_t = a_t'$, $a_{t-1} = a_{t-1}', \dots, a_{t-l} = a_{t-l}'$, and similarly for $\bar{m}, \bar{m}'$. Then we have
\begin{align*}
Y_{it}(\bar{a}, \bar{m}) = Y_{it}(\bar{a}', \bar{m}'), \qquad \forall \bar{a}, \bar{m}, \bar{a}', \bar{m}'.
\end{align*}
\end{assumption}

\noindent This extends the previous l-limited carryover effects assumption to incorporate multiple policies. The key difference is that carryover effects are limited to the previous $l$ time-periods for both policies.

\begin{assumption}[Ignorability]\label{asmpt-3a}
\begin{align*}
Y_{it}(\bar{a}, \bar{m}) \perp (A_{it}, M_{it}) \mid \bar{Y}_{it-1}(\bar{0}, \bar{0}), \qquad \forall \bar{a}, \bar{m}
\end{align*}
\end{assumption}

\noindent The multiple policy ignorability assumption states that both policies are jointly randomized with respect to the potential outcomes given the potential outcomes absent treatment, and allows for causal pathways between the two policies, so that $M$ ($A$) may lie on the pathway from $A$ ($M$) to $Y$. We next assume constant treatment effects,

\begin{assumption}[Constant treatment effects]\label{asmpt-4a}
\begin{align*}
Y_{it}(\bar{a}, \bar{m}) = Y_{it}(0, 0) + \theta_t(\bar{a}, \bar{m})
\end{align*}
\end{assumption}

This assumption again precludes effect heterogeneity on the additive scale (though this could be modified to the multiplicative scale if desired).  

\begin{assumption}[Positivity]\label{asmpt-5a}
For some $\epsilon > 0$ and $e(y, \bar{a}, \bar{m}) = \P\left\{\bar{A}_{it}=\bar{a}, \bar{M}_{it}=\bar{m} \mid Y_{i,t-1}(\bar{0}) = y\right\}$, we have
\begin{align*}
\min_{\bar{a}, \bar{m}} P\left[e\{Y_{i,t-1}(0), \bar{a}, \bar{m}\} \ge \epsilon\right] = 1.
\end{align*}
\end{assumption}

\noindent This assumption again simply extends the previous positivity assumption by stating that there is some positive probability of observing any treatment history value for either treatment conditional on a specific value of the previous potential outcome absent any treatments. Finally, we extend assumption \ref{asmpt-linear}:

\begin{assumption}[Parametric model]\label{asmpt-6a}
Assume that for each unit $i$ and time period $t$,
\begin{align}
\label{eqnres1}Y_{it}(\bar{0}, \bar{0}) = \alpha_t +\beta_t^\top\phi(Y_{i,t-1}(\bar{0}, \bar{0})) + \epsilon_{it}.
\end{align}
\noindent for some parameter vector $(\alpha_1, \beta_1^\top, \dots, \alpha_T, \beta_T)^\top \in \mathbb{R}^{T(q+1)}$ and some known function $\phi(x): \mathbb{R}\to\mathbb{R}^q$. Let $Z_{it} = [1, A_{it}, \phi^\top(Y_{i,t-1}(\bar{0}, \bar{0}))]$ and assume that $\E[Z_{it}^\top Z_{it}]$ has full column rank.
\end{assumption}

\begin{proposition}\label{prop:para-extension}
Under assumptions \ref{asmpt-1a}-\ref{asmpt-6a}, $\mathbb{E}[Y_{it}(\bar{a}, \bar{m})] - \mathbb{E}[Y_{it}(\bar{a}', \bar{m}')]$ is identified in the data as,
\begin{align*}
 \theta_a^{\star\top} (\bar{a} - \bar{a}') + \theta_m^{\star\top} (\bar{m} - \bar{m}'),
\end{align*}

\noindent where $(\theta_a^\star, \theta_m^\star)$ are minimizers of $\E[(Y_{it} - \beta_t-\theta_a^\top\bar{A} - \theta_m^\top\bar{M} - \beta_t^\top\phi(Y_{it-1} - \theta_a^\top\bar{A}_{i,t-1}-\theta_m^\top\bar{M}_{i,t-1})^2]$.    
\end{proposition}

\section{More details on differential measurement error}
In this section, we allow the measurement error to depend on any of the terms in the model, and highlight the bias obtained in this more general scenario than what was considered in the manuscript. This results in a more general expression of the bias compared to the bias expression seen in Section \ref{sec:relaxations}. Specifically, we can show that our estimate of the treatment effect in the presence of measurement error induced by treatment effect heterogeneity converges to
\begin{align*}
    \frac{\text{cov}(A_{it},Y_{it}) - \frac{\text{cov}(A_{it},Y_{i,t-1}(0) + U_{i,t-1}) \text{cov}(Y_{it},Y_{i,t-1}(0) + U_{i,t-1})}{\text{var}(Y_{i,t-1}(0) + U_{i,t-1})}}{\text{var}(A_{it}) - \frac{\text{cov}^2(A_{it},Y_{i,t-1}(0) + U_{i,t-1})}{\text{var}(Y_{i,t-1}(0) + U_{i,t-1})}}.
\end{align*}

\noindent If the measurement error is classical, then any covariance terms involving $U_{i,t-1}$ are zero and therefore the measurement error only impacts the $\text{var}(Y_{i,t-1}(0) + U_{i,t-1})$ term, which simplifies the formula to what was seen in equation \eqref{eqn:ClassicalFormula} of the main manuscript. If, however, the measurement error is differential, the bias can take different magnitudes or directions. Differential measurement error here could occur if $U_{i,t-1}$ is correlated with any of $A_{it}, Y_{it},$ or $Y_{i,t-1}(0)$. It is thus worth considering in our context whether we expect the bias to be differential, if we have any prior knowledge about the direction of bias, and whether it depends on the source of heterogeneity. 

We discussed the impact of correlation between $U_{i,t-1}$ and $Y_{it}$ in the manuscript and the possible bias that can occur in that scenario. Here, we can discuss differential measurement error caused by correlation between $U_{i,t-1}$ and either $A_{it}$ or $Y_{it}(0)$. We believe it is unlikely to have dependence between $A_{it}$ and $U_{i,t-1}$ as units are not likely to be treated based on the magnitude of the individual level treatment effect. If this were to occur, however, it would likely stem from positive dependence between $A_{it}$ and $U_{i,t-1}$, the implications of which would be essentially identical to what was discussed in the manuscript with differential measurement error caused by dependence between $Y_{it}$ and $U_{i,t-1}$. This would lead to treatment effect estimates that are less than what would be seen for the same magnitude of measurement error if it were instead classical. 

The last form of differential measurement error to consider is the case where $U_{i,t-1}$ is correlated with $Y_{i,t-1}(0)$. This could occur if units that have larger values of $Y_{it}(0)$ are more likely to benefit from treatment. In the opioid setting, states with higher mortality from opioid use might be more likely to benefit from policies reducing its impact. One thing to note is that dependence between $U_{i,t-1}$ and $Y_{i,t-1}(0)$ makes it very likely to have dependence between $U_{i,t-1}$ and $Y_{it}$ as well given the clear relationship between potential outcomes and observed outcomes. For this reason, many of the issues mentioned in the manuscript for dependence with $Y_{it}$ would also apply in this situation; however, here we focus specifically on the impacts of the dependence between $U_{i,t-1}$ and $Y_{i,t-1}(0)$. Letting $P_y(U_{i,t-1})$ denote the projection of $U_{i,t-1}$ onto $Y_{i,t-1}(0)$, the variable we adjust for can be written as
\begin{align*}
    Y_{i,t-1}(0) + U_{i,t-1} &= Y_{i,t-1}(0) + P_y(U_{i,t-1}) + U_{i,t-1} - P_y(U_{i,t-1}) \\
    &= Y_{i,t-1}^*(0) + U_{i,t-1}^*,
\end{align*}

\noindent where $Y_{i,t-1}^*(0) = k Y_{i,t-1}(0)$ for some constant $k$, and $U_{i,t-1}^* = U_{i,t-1} - P_y(U_{i,t-1})$. Importantly, we have that
$\text{Cov}(Y_{i,t-1}^*(0), U_{i,t-1}^*) = 0$, which means that differential measurement error with respect to $Y_{i,t-1}(0)$ can be equivalently thought of as measurement error that is not correlated with $Y_{i,t-1}^\star(0)$ and the degree of error is driven by the magnitude of $U_{i,t-1}^*$. The discussion on classical measurement error therefore holds in this setting; however, importantly, $\text{Var}(U_{i,t-1}^*) \leq \text{Var}(U_{i,t-1})$. The bias should therefore be smaller in this setting relative to error of the same magnitude that isn't correlated with $Y_{i,t-1}(0)$. 

\section{Inference on sample level estimands}

In our simulation studies and data analysis, we used our model to estimate quantities other than population average treatment effects. This is particularly useful for comparing our approach to existing estimators such as synthetic controls and DiD estimators, which commonly target average treatment effects on the treated over specific time periods for each unit. For example, one frequently encountered estimand is a sample average treatment effect on the treated. This is defined as,

\begin{align*}
SATT_t = \frac{1}{n_t} \sum_{i \in \mathcal{N}_t} Y_{it}(1) - Y_{it}(0),
\end{align*}

\noindent where $\mathcal{N}_t$ is the set of indices corresponding to the treated units and $n_t$ is the number of treated units at time $t$ (note here that we assume $l=0$ and therefore the potential outcome at time $t$ only depends on treatment at time $t$). 

In principle it is straightforward to target sample estimands with our model; however, inference requires additional consideration. For more information on distinctions between sample and population estimands, as well as Bayesian inference for them, see \cite{ding2018causal} and \cite{li2023bayesian}. We consider two natural approaches to inference in this setting that both require calculating the posterior distribution of 
\begin{align*}
\frac{1}{n_t} \sum_{i \in \mathcal{N}_t} Y_{it}^*(1) - Y_{it}^*(0).
\end{align*}
\noindent These approaches differ, however, in defining the posterior distribution for $Y_{it}^*(a)$. The first approach sets $Y_{it}^*(A_{it}) = Y_{it}$ and draws $Y_{it}^*(1 - A_{it})$ from an appropriate distribution (for example, normal or negative binomial) with parameters given by our model. This approach acknowledges that one of the counterfactuals is observed and does not need to be imputed, and then imputes the other from the model. The second approach draws both $Y_{it}^*(1)$ and $Y_{it}^*(0)$ from the posterior predictive distribution; for example, a normal or negative binomial distribution with parameters given by our model. This approach ignores information from the observed counterfactual and instead imputes both assuming they are independent. The first approach may seem more natural as it uses the observed potential outcome. In our experience, however, this approach works well when the model is correctly specified, but suffers more from moderate amounts of model misspecification relative to the second approach. Many components of the model may be misspecified, including the treatment effect function. If components other than the treatment effect portion of the model are misspecified, we will obtain poor predictions of $Y_{it}^*(1 - A_{it})$, in turn leading to poor estimates of $SATT_t$. The second approach instead uses the difference in two posterior predictive draws, $Y_{it}^*(1) - Y_{it}^*(0)$: this difference may better estimate the treatment effect, even when individually each of the draws $Y_{it}^*(a)$ are biased for the true potential outcome. While both approaches are reliant to some degree on correct model specification, the first approach relies on estimating the full data generating process well, while the second is slightly less reliant on the entire model, relying more on the treatment effect specification itself.

\section{Additional information and results for simulation study}

\subsection{Data generation mechanism for simulation study}

We describe in more detail the specifics of the simulation study used in the manuscript. To vary the degree of confounding, we assign treatment to units with higher values of the outcome, and allow treatment times to vary as a function of the outcome as well. We first calculate the outcome for each unit in time period 7, and then used max-min scaling to let all these values fall between 0 and 1. We refer to this value as $\widetilde{Y}_7$. To generate start times that better reflects staggered adoption, we randomly sample $20$ numbers between 8 and 14 with replacement and sort them in increasing order. We can refer to this ordered vector of potential times as $\widetilde{T}_1, \dots, \widetilde{T}_{20}$ where we have that $\widetilde{T}_1 \leq \widetilde{T}_2 \leq \dots \leq \widetilde{T}_{20}$. We assign the $j^{th}$ time $\widetilde{T}_j$ to be the start time for unit $i$ with probability proportional to $\widetilde{Y}_{7,i}^c$, where $c$ is what we refer to as the confounding constant. Once a unit has been assigned a start time, they are removed from the pool of units moving forward and we continue until 20 units have a treatment time. This process ensures that units with higher values of the outcome are more likely to initiate treatment earlier, and the units that are treated have higher outcome values generally. When $c=0$ there is no confounding by $Y_{it}(0)$ and units and times are chosen completely randomly, but as $c$ increases, the degree of confounding increases. To facilitate comparison, we also ensure that units remain treated for all time periods after treatment initiation, though this is not required for our approach.

We also need to generate a treatment effect for each unit in the sample. To ensure that larger states have larger treatment effects, we first work on the rate scale, which is the number of outcomes divided by the population size of each state. We assume the effect is homogeneous with respect to time on the rate scale, but is unique to each state. Formally, we generate new rates $R_{it}^*$ as a function of the observed rates $R_{it}$ via
\begin{align*}
R_{it}^* = R_{it} + A_{it}(\theta + \theta_i).
\end{align*}
\noindent Throughout, we set $\theta = -0.02$ to fix a small average impact of the policy on rates of prescribing, and we vary $\theta_i$ depending on what level of heterogeneity we are considering. We set $\theta_i = k \boldsymbol{X}_i \boldsymbol{\gamma}$ where $\boldsymbol{X}_i$ are time-invariant characteristics, and the elements of $\boldsymbol{\gamma}$ are generated from independent standard normal distributions. We explore $k \in \{0, 0.009, 0.026 \}$, which we denote by no heterogeneity, moderate heterogeneity, and high heterogeneity. The two non-zero values were chosen to ensure that $P(|\theta_i| > |\theta|)$ was equal to 0.1 or 0.33, respectively. This implies that in the high heterogeneity setting, approximately one third of the subjects have unit-specific treatment effects that are either of the opposite sign of the average treatment effect, or are at least twice as large. The observed outcomes are set as $Y_{it}^* = R_{it}^* P_{it}$, where $P_{it}$ is the population size for unit $i$ at time $t$, and outcomes are rounded to the nearest integer to ensure count outcomes. 

\subsection{Negative binomial simulations on opioid prescribing}

We evaluate the performance of the proposed autoregressive approach when using negative binomial models. As in the simulation section of the manuscript, we utilize the observed opioid data to perform simulation studies instead of using fully synthetic data. We follow the same simulation structure as in Section \ref{sec:Simulations} of the manuscript that allows for staggered adoption of the policy and confounding by previous time points. The only difference between the data generation in the manuscript and in this section is how the treatment effects are generated. As in the manuscript, a small average treatment effect is incorporated, and we vary the degree of heterogeneity of the treatment effect across units. We choose the degree of heterogeneity in the same manner as in the manuscript, but we generate new rates on the multiplicative scale via:
$$\frac{R_{it}^*}{R_{it}} = \exp \big( A_{it}(\theta + \theta_i) \big),$$
so that Assumption \ref{assume:ConstantMultiplicative} is satisfied when there is no heterogeneity ($\theta_i = 0$ for all $i$). As in the manuscript, the constant treatment effects assumption will be violated in the moderate and high heterogeneity situations. We target the same estimand as in Section \ref{sec:Simulations}, given by
\begin{align*}
\frac{1}{n_t} \sum_{i \in \mathcal{N}_t} \sum_{m=0}^2 \bigg\{ Y_{i, T_{i} + m}(1) - Y_{i, T_{i} + m}(0) \bigg\},
\end{align*}

\noindent which targets the average treatment effect on the treated units in the first three time periods after policy adoption. We compare our approach using negative binomial autoregressive models as described in Section \ref{ssec:Nonlinear} of the manuscript with the difference in differences, synthetic control, and two-way fixed effects approaches described in Section \ref{sec:Simulations}. 

\begin{figure}[H]
    \centering
    \includegraphics[width=0.85\linewidth]{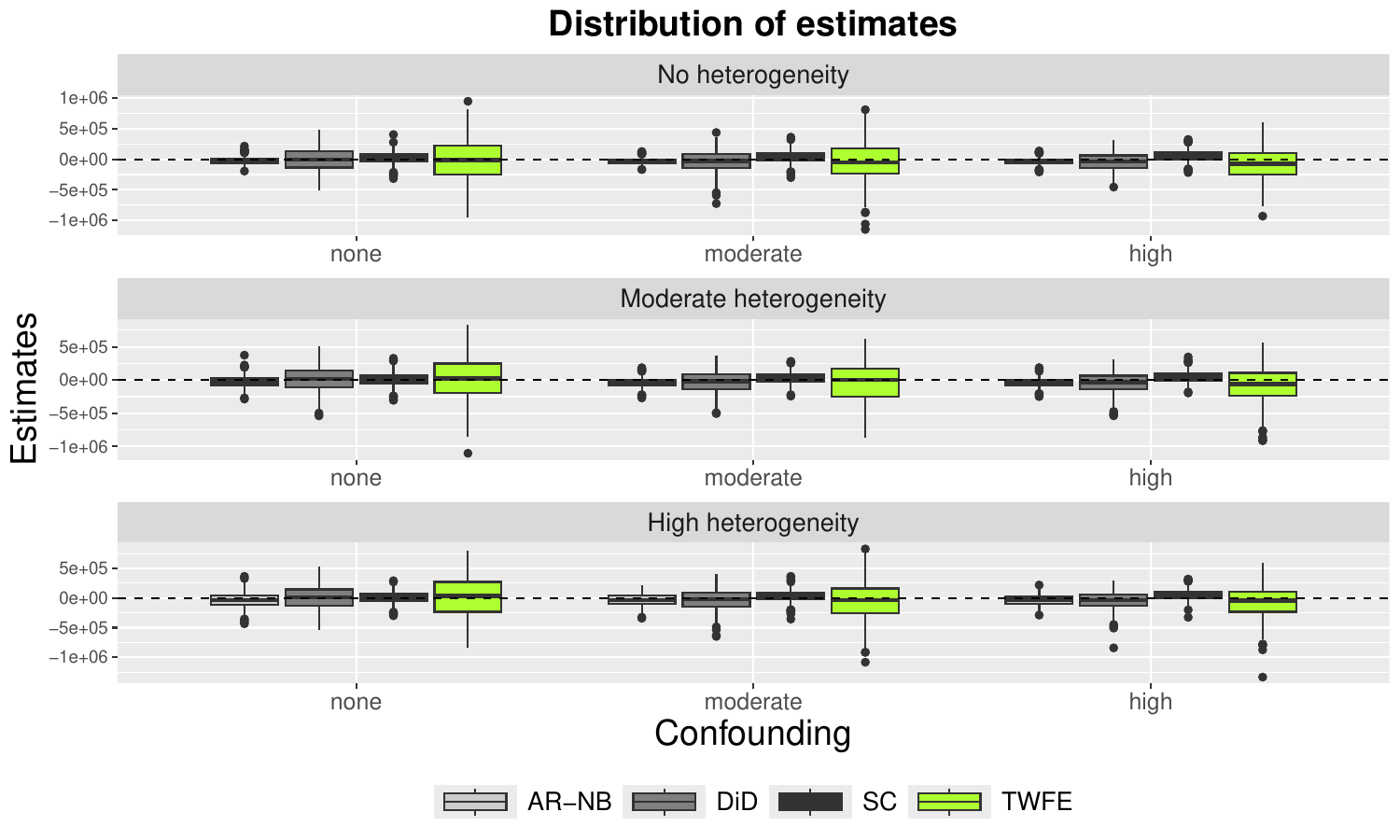}
    \caption{Distribution of point estimates for simulations on opioid prescribing data with negative binomial autoregressive models.}
    \label{fig:23results}
\end{figure}

\begin{figure}[H]
    \centering
    \includegraphics[width=0.85\linewidth]{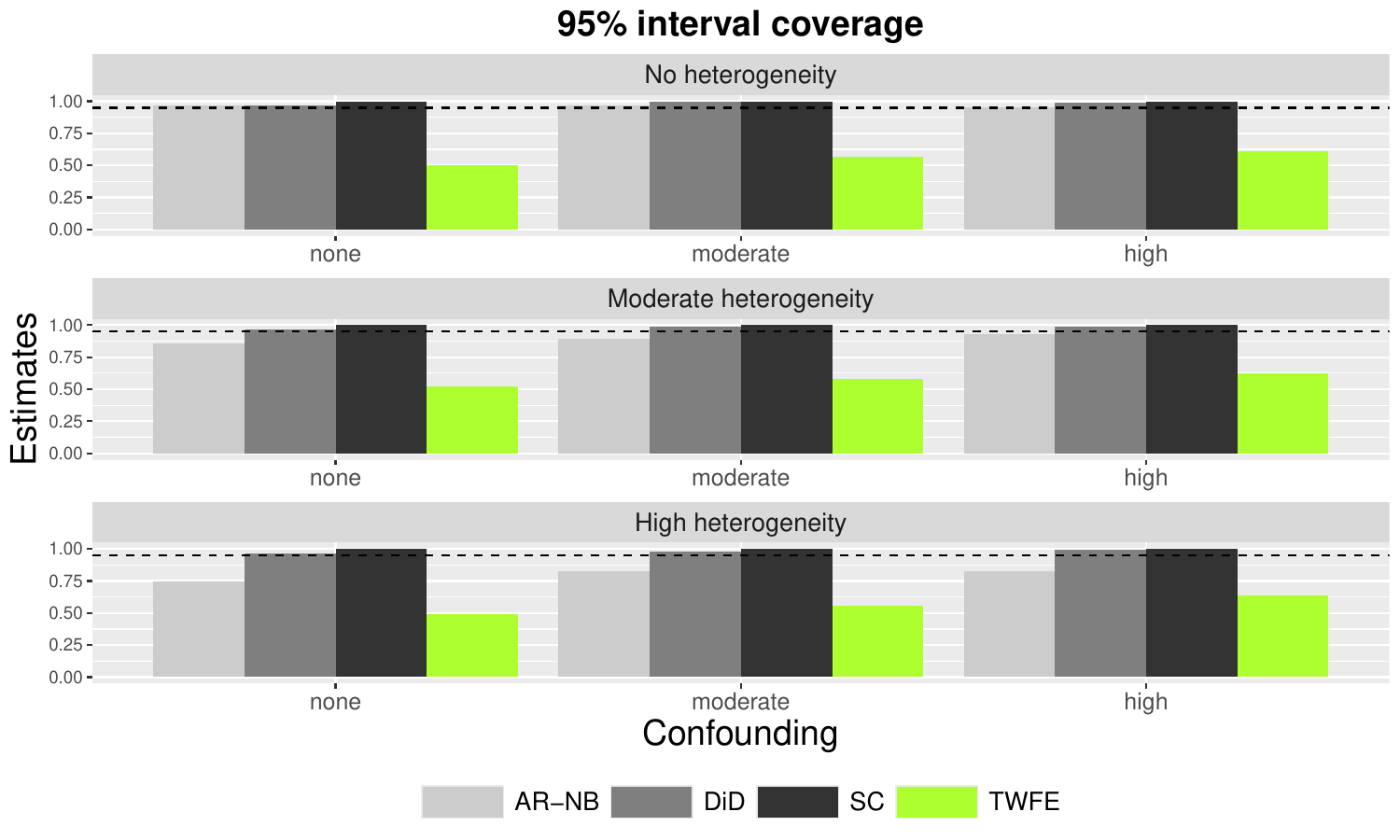}
    \caption{95\% interval coverage for simulations on opioid prescribing data with negative binomial autoregressive models.}
    \label{fig:23resultsCoverage}
\end{figure}

\begin{figure}[H]
    \centering
    \includegraphics[width=0.85\linewidth]{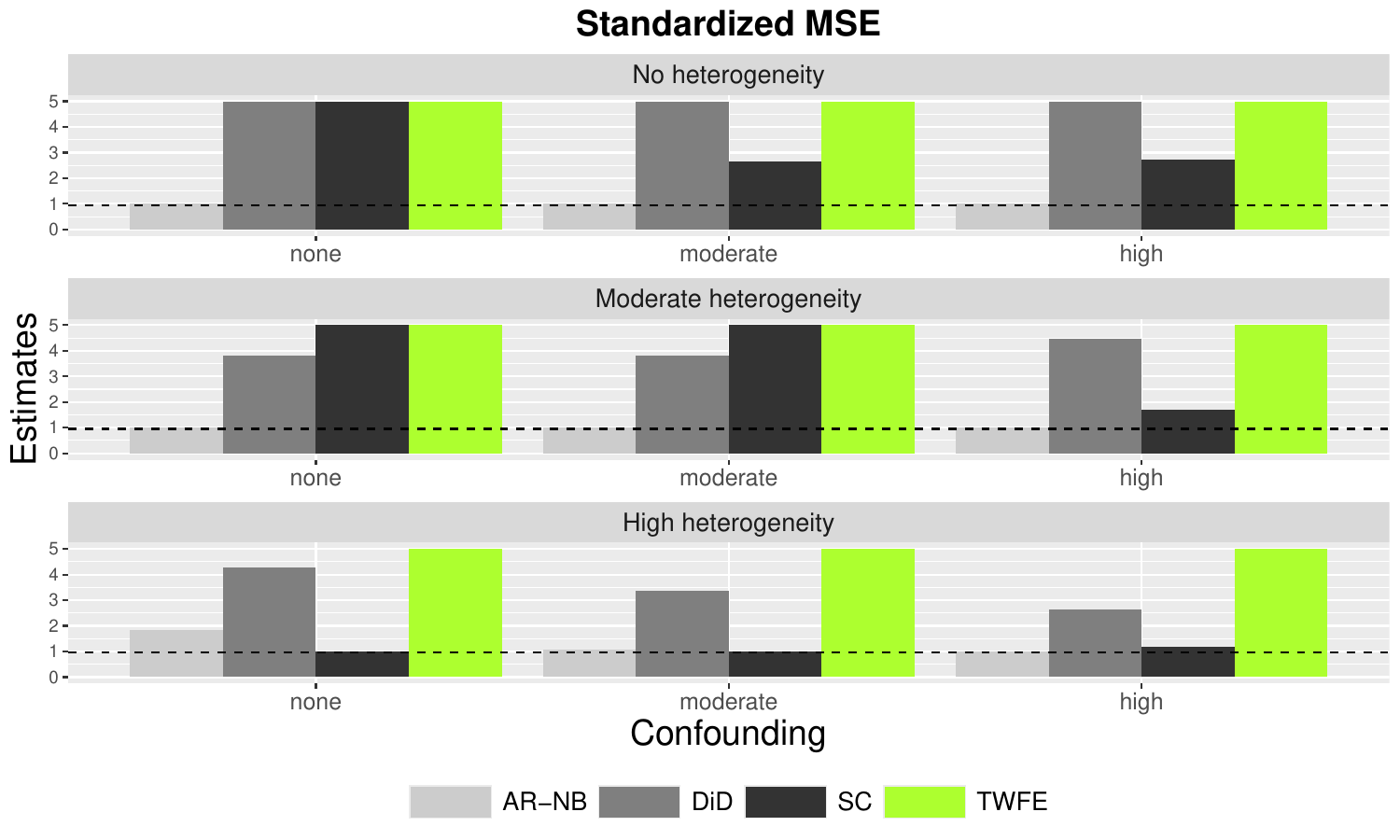}
    \caption{Mean squared error for simulations on opioid prescribing data with negative binomial autoregressive models. MSE values have been capped at 5 for easier comparison.}
    \label{fig:23resultsMSE}
\end{figure}

\begin{figure}[H]
    \centering
    \includegraphics[width=0.85\linewidth]{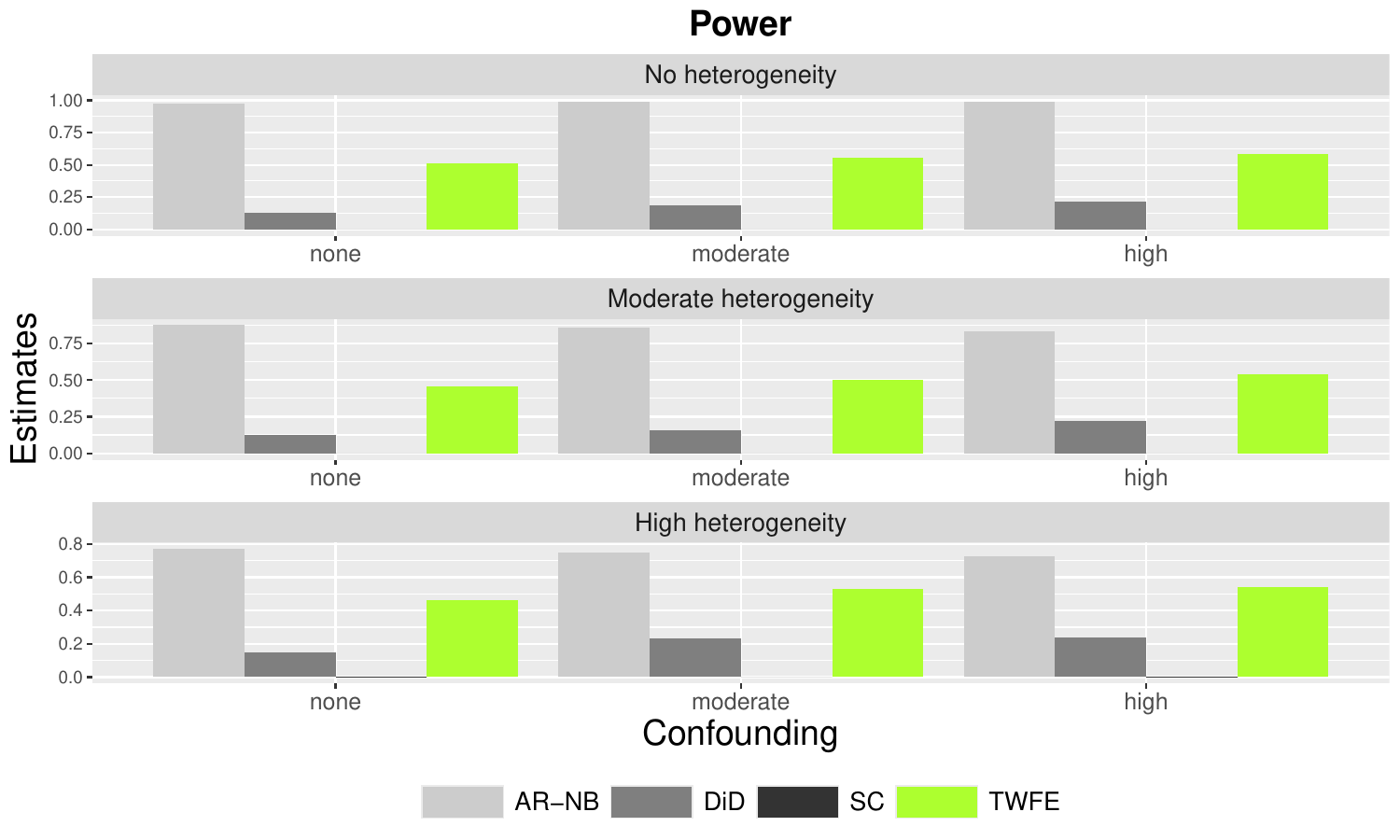}
    \caption{Power for simulations on opioid prescribing data with negative binomial autoregressive models.}
    \label{fig:23resultsPower}
\end{figure}

The results can be found in Figures \ref{fig:23results} - \ref{fig:23resultsPower}. Overall, the results mirror those seen in Section \ref{sec:Simulations} when using the linear autoregressive models rather than the negative binomial autoregressive models. The improved efficiency leads to large decreases in MSE compared with existing estimators. The synthetic control estimator performs slightly better in terms of MSE in some of the high heterogeneity simulations, though the autoregressive model has the lowest MSE in all other scenarios. The interval coverage for the autoregressive model is near the nominal rate, except in situations with very large degrees of heterogeneity, which echoes the same findings from Section \ref{sec:Simulations} of the manuscript. Lastly, the autoregressive approach has the highest power of all four of the approaches to identify a treatment effect across all simulation scenarios, due to the improved efficiency of the approach. 

\section{Additional information and results for opioid analysis}\label{app:E}

\begin{figure}[htbp]
    \centering
    \includegraphics[width=0.6\textwidth]{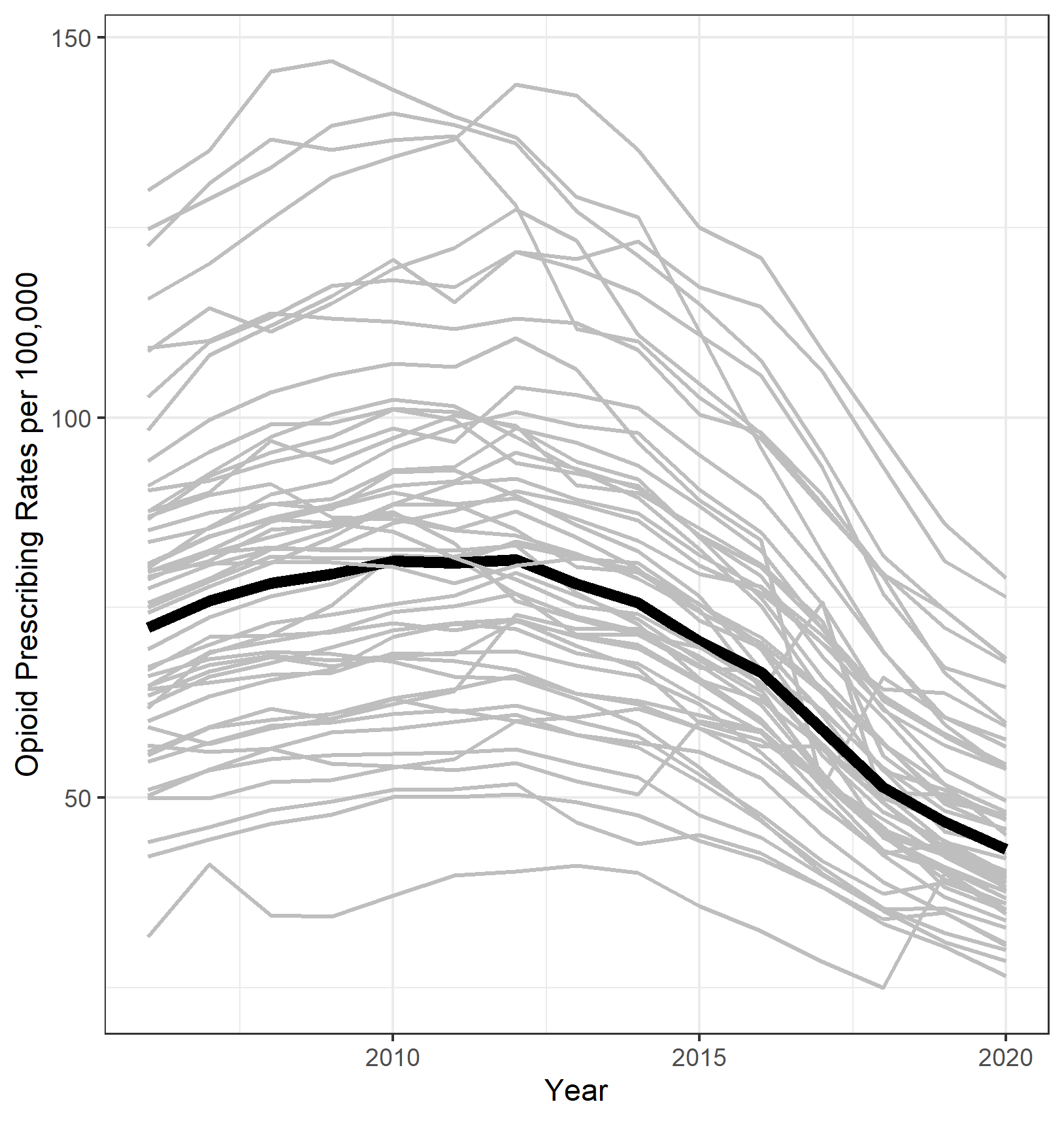}
    \caption{Trend of Opioid Prescribing Rates per 100,000 people, separated by state. The dark line represents the national average. }
    \label{fig:rxper100}
\end{figure}

\begin{figure}[htbp]
    \centering
        \includegraphics[width=0.8\textwidth]{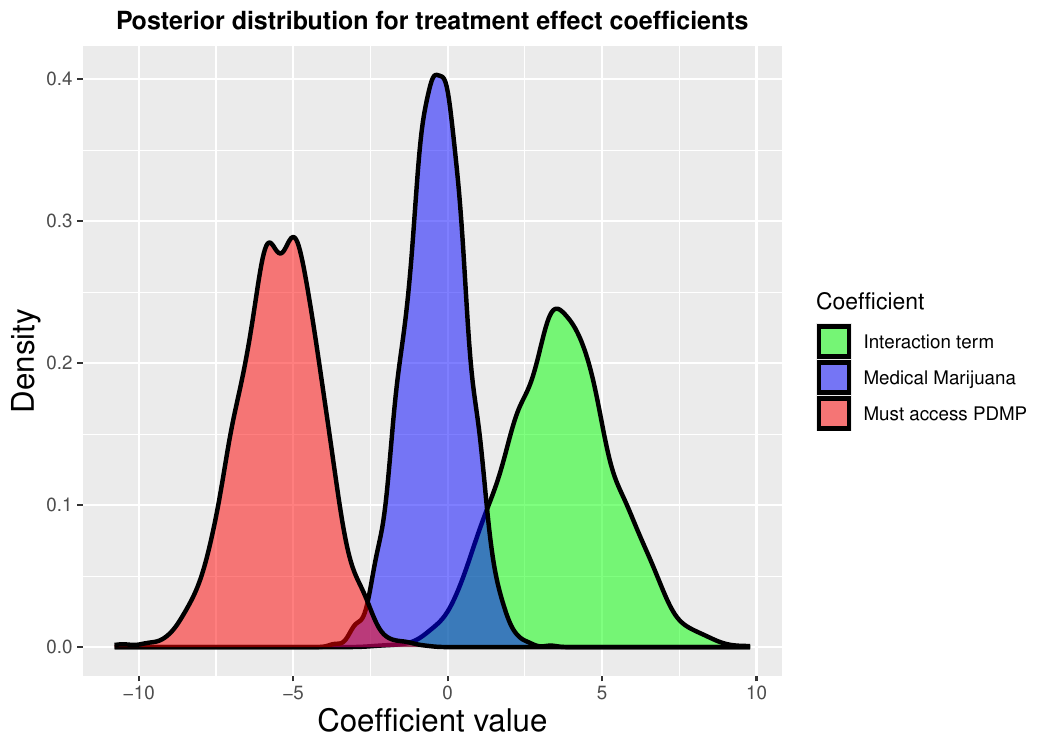}
    \caption{Posterior distribution of treatment effect parameters in the autoregressive model for the analysis of PDMP and medical marijuana laws jointly. }
    \label{fig:MultiplePolicyPosterior}
\end{figure}

We provide additional information about the data used in Section \ref{sec:Application}, as well as some additional results that were not shown in the manuscript. Figure \ref{fig:rxper100} shows time trends for the rate of opioid prescriptions by state for each year in the study, with the solid black line representing the nationwide average. We see that in the early years of the study opioid prescriptions were rising, though these began to fall in the second half of the study. Figure \ref{fig:MultiplePolicyPosterior} shows the distribution of the model parameters corresponding to the treatment effect in the multiple policy setting. The red histogram shows that the effect of must access PDMP laws reduces the rate of opioid prescriptions, the blue histogram shows that medical marijuana has essentially no effect on prescription rates, and the interaction term in green shows that the effect of must access PDMP laws is indeed modified by the presence of medical marijuana laws.

\section{Proofs}

In this section, we prove propositions \ref{prop:para-ident}, \ref{prop:nonpara-ident}, and \ref{prop:para-extension}.

\begin{proof}[Proof of proposition \ref{prop:para-ident}]
\begin{align*}
Y_{it}(0) &= \alpha_t + \beta_t^\top\phi(Y_{it-1}(0)) + \epsilon_{it} \\
&= \alpha_t + \beta_t^\top\phi(Y_{it-1}-\theta A_{it-1}) + \epsilon_{it} 
\end{align*}
\noindent where the first line follows from assumption \ref{asmpt-linear}, and the second from assumption \ref{asmpt-uniform} and \ref{asmpt-consistency}. Then, again applying \ref{asmpt-uniform} and \ref{asmpt-consistency}, we see that

\begin{align*}
Y_{it} = \alpha_t + \theta A_{it} + \beta_t^\top(Y_{it-1}-\theta A_{it-1}) + \epsilon_{it},
\end{align*}

\noindent where $\E[\epsilon_{it} \mid Y_{it-1}-\theta A_{it-1}, A_{it}] = 0$ by ignorability and consistency. Because $Z'Z$ is invertible, where $Z_{it} = [1, A_{it}, Y_{it-1}(0)] = [1, A_{it}, Y_{it-1}-\theta A_{it-1}]$, we may conclude that $\theta$ is identified.
\end{proof}

\begin{proof}[Proof of proposition \ref{prop:nonpara-ident}]
Note that

\begin{align*}
\E[Y_{i,t^\star+1}(1)] &= \E[\E[Y_{i,t^\star+1}(1) \mid Y_{it^\star}(0)]] \\
&= \E[\E[Y_{i,t^\star+1}(1) \mid Y_{it^\star}(0), A_{i,t^\star+1} = 1]] \\
&= \E[\E[Y_{i,t^\star+1} \mid Y_{it^\star}, A_{i,t^\star+1} = 1]],
\end{align*}

\noindent where the first equality follows by iterating expectations, the second by ignorability, and the final assumption by consistency and the assumption that $A_{it^\star} = 0$ for all units at time $t^\star$. This exact same logic holds setting $a = 0$, and thus $\theta$ is identified.
\end{proof}

\begin{proof}[Proof of proposition \ref{prop:para-extension}]
This proof trivially follows from the proof of \ref{prop:para-ident}, where we treat $(A_{it}, M_{it})$ as a multivariate treatment, and simply substitute this into the derivation above.
\end{proof}

\section{Bias bound under the linear factor model}\label{app:bound}

In this section we examine the bias of the linear auto-regressive model under a linear factor model. Specifically, we consider the case where the potential outcomes are generated by a factor model plus a constant treatment effect:
\begin{align*}
Y_{it}(0) &= \delta_t + \blambda\bmu + \epsilon_{it}, \\
Y_{it}(1) &= Y_{it}(0) + \theta,  
\end{align*}
\noindent so that\
\begin{align}
\label{eqn1} Y_{it} &= \delta_t + \theta A_{it} + \blambda\bmu + \epsilon_{it}, 
\end{align}
\noindent where $\blambda \in \mathbb{R}^{1 \times r}, \bmu \in \mathbb{R}^{r \times 1}$. We also assume that the errors $\epsilon_{it}$ are,

\begin{enumerate}
    \item Mean-zero: $\E_t[\epsilon_{it} \mid \bmu] = 0$
    \item Finite-variance: $Var_t[\epsilon_{it}\mid \bmu] \le \sigma^2$ for all $i, t$
    \item Uncorrelated across time: $\epsilon_{it} \perp \epsilon_{is} \mid \bmu, \qquad\forall t \ne s$
\end{enumerate}

We further assume that treatment $A_{it}$ is random conditional on $\bmu$ (i.e. $A_{it} \perp Y_{is}(a) \mid \bmu$, for all $s, t$), where $A_{it} \sim \text{Bern}(g_t(\bmu))$ for some function $g_t$, and that $\P(1 - \eta \ge g_t(\bmu) \ge \eta) = 1$ for all $t$ and some $\eta > 0$. Let $r \ge 1$ and define the $m$ by $r$ matrix of the $m$ most recent factor loadings at time $t$, $\bLambda$, which we treat as fixed. We similarly define the vector of the $m$ most recent lagged untreated outcomes $\bY(0)$, time fixed-effects $\bdelta$, and errors $\bepsilon$. Finally, assume that the minimum eigenvalue of $\frac{1}{m}\bLambda^\top \bLambda$ is bounded away from zero by $\underline{\zeta}$ for all $t$ and $m$, and that $\|\blambda\|\le\bar{\lambda}$ for all $t$, where $\|A\|$ denotes the L2 norm for a vector $A$, and the operator norm for a matrix $A$.

\begin{proposition}\label{prop:2}
Let $\tilde\theta_m$ be the minimizer of $\E[(Y_{it} - \alpha_t - \theta A_{it} - \beta_{t}^\top \bY(0))^2]$, where $\bY(0)$ has length $m$. Under the assumptions stated above, $\lvert\tilde{\theta}_m - \theta\rvert \le Cm^{-1/2}$ for some constant $C$. 
\end{proposition}

\begin{proof}
\begin{align}
\nonumber Y_{it} &= \delta_t +\theta A_{it} + \blambda\bmu + \epsilon_{it} \\
\nonumber&= \delta_t + \theta A_{it} + \blambda (\bLambda^\top\bLambda)^{-1}\bLambda^\top(\bY(0) - \bdelta- \bepsilon) + \epsilon_{it} \\
\nonumber&= \alpha_t  + \theta A_{it} + \beta_t^\top\bY(0) - \blambda(\bLambda^\top\bLambda)^{-1}\bLambda^\top\bepsilon + \epsilon_{it} \\
\label{eqn3}&= \alpha_t + \theta A_{it} + \beta_t^\top\bY(0) + v_{it} + \epsilon_{it}, 
\end{align}

\noindent Moreover, for any time-period $t$,

\begin{align*}
\E_t [v_{it}] = -\E_t [\blambda(\bLambda^\top\bLambda)^{-1}\bLambda^\top\bepsilon] = \blambda(\bLambda^\top\bLambda)^{-1}\bLambda^\top\E_t[\bepsilon] = 0,
\end{align*}

\noindent and

\begin{align*}
\E_t [v_{it}^2] &= \blambda(\bLambda^\top\bLambda)^{-1}\bLambda^\top\bSigma \bLambda (\bLambda^\top\bLambda)^{-1}\blambda^\top,
\end{align*}

\noindent where $\bSigma = \text{diag}(\sigma_{i,t-1}^2, ..., \sigma_{i,t-m}^2)$ due to the assumed independence of $\epsilon_{it}$ over time. Then,

\begin{align*}
\E_t [v_{it}^2] &\le \|\blambda\|^2\|(\bLambda^\top\bLambda)^{-1}\|^2\|\bLambda^\top\bSigma\bLambda\| \\
&\le \left(\frac{\bar{\lambda}}{m\underline{\zeta}}\right)^2 \|\sum_{s=1}^m \boldsymbol\lambda_{t-s}^\top\sigma_{i,t-s}^2\boldsymbol\lambda_{t-s}\| \\
&\le \left(\frac{\bar{\lambda}^2}{\underline{\zeta}}\right)^2\frac{\sigma^2}{m}.
\end{align*}

\noindent Let $\ddot W_{it} = W_{it} - \E_t[W_{it}]$ for some random variable $W_{it}$. Partialling out the time fixed-effects, we may re-write equation (\ref{eqn3}) as,

\begin{align}
\label{eqn4}\ddot Y_{it} = \theta \ddot A_{it} + \beta_t^\top \dbY + v_{it} + \epsilon_{it}.
\end{align}

\noindent Define $\tilde A_{it} = \ddot A_{it} - \pi_t^\top \dbY(0)$, where $\pi_t = \arg \min_\pi \E_t[(\ddot A_{it} - \pi^\top \dbY(0))^2]$. First, notice that, 

\begin{align*}
Cov_t(\tilde A_{it}, \epsilon_{it}) &= Cov_t(\ddot A_{it} -\pi_t^\top\dbY(0),\epsilon_{it}) \\
&= \E_t[\ddot A_{it}\epsilon_{it}] - \pi_t^\top \E_t[\ddot A_{it}\dbY(0)] \\
&= \E_t[\E_t[\ddot A_{it}\epsilon_{it} \mid \bmu]] - \pi_t^\top \E_t[\E_t[\ddot A_{it}\dbY(0) \mid \bmu]] \\
&= - \pi_t^\top\E_t[\E_t[\ddot A_{it}\bepsilon\mid \bmu]] = 0. \\
\end{align*}

\noindent Noting that the $\theta^\star$ that minimizes $\E[(\ddot Y_{it} - \theta \ddot A_{it} + \beta_t^\top \dbY(0))^2]$ is equal to $\theta_m$, we may conclude that,

\begin{align*}
\theta_m &= \theta + \frac{Cov(\tilde A_{it}, v_{it})}{Var(\tilde{A}_{it})} = \theta+ \frac{\sum_tCov_t(\tilde A_{it}, v_{it})}{\sum_tVar_t(\tilde A_{it})}, \\
\implies \lvert\tilde{\theta}_m - \theta\rvert &\le C_\eta^2 \max_t \lvert Cov_t(\tilde A_{it}, v_{it})\rvert\le C_\eta \sqrt{Var_t(v_{it})} \\
&\le C_\eta\left(\frac{\bar{\lambda}^2}{\underline{\zeta}}\right)\frac{\sigma}{\sqrt{m}}.
\end{align*}

\noindent where $C_\eta = \frac{1}{\sqrt{\eta(1-\eta)}}$, and noting that $Var_t(\tilde A_{it}) = \E_t[(\ddot A_{it} - \pi_t^\top\dbY(0))^2] \ge \E_t[(\ddot A_{it} - \E_t[\ddot A_{it} \mid \bmu])^2] \ge \eta(1-\eta)$.
\end{proof}

\begin{remark} This shows that the bias of the estimator of $\theta$ decreases as we increase the number of lagged potential outcomes absent treatment; moreover, if we allow $m = \alpha T$ for $1 \ge \alpha > 0$, this further implies that the bias is $\mathcal{O}(T^{-1/2})$. This contrasts with the result from \cite{abadie2010synthetic}, where the bias decreases with $T_0$, the number of pre-treatment outcomes.
\end{remark}

\begin{remark}Our proposed estimation procedure differs from this proof in that we estimate a common coefficient vector on the lagged potential outcomes. This strategy is justifiable when the propensity score model is a constant function of the pre-period outcomes over time, however, will result in bias if this does not hold. To see this, notice that when we impose a constant coefficient on the pre-period potential outcomes we obtain,

\begin{align*}
\ddot Y_{it} &= \theta \ddot A_{it} + \beta^\top \dbY(0) + (\beta_t - \beta)^\top\dbY(0) + v_{it} + \epsilon_{it} \\
&= \theta \ddot A_{it} + \beta^\top \dbY(0) + u_{it}.
\end{align*}

\noindent Now re-define $\tilde A_{it}$ as,

\begin{align*}
\tilde A_{it} &= \ddot{A_{it}} - \pi_t^\top \dbY(0) + (\pi_t - \pi)^\top \dbY(0) \\
&= r_{it} + (\pi_t - \pi)^\top \dbY(0),
\end{align*}

\noindent where $\pi = \arg\min_\pi \E[(\ddot A_{it} - \pi^\top \dbY(0))^2]$. Then for each time period $t$ we have that,

\begin{align*}
Cov_t(\tilde A_{it}, u_{it}) &= Cov_t(r_{it} + (\pi_t - \pi)^\top \dbY(0), (\beta_t - \beta)^\top\dbY(0) + v_{it} + \epsilon_{it}) \\
&= (\beta_t - \beta)^\top Cov_t(r_{it}, \dbY(0)) + (\pi_t - \pi)^\top \Sigma_{\dbYn(0),t} (\beta_t - \beta) + Cov_t(\tilde A_{it}, v_{it}) + Cov_t(\tilde A_{it},  \epsilon_{it}) \\
&= (\pi_t - \pi)^\top \Sigma_{\dbYn(0), t} (\beta_t - \beta) + \mathcal O(m^{-1/2}),
\end{align*}

\noindent where $\Sigma_{\dbYn(0), t} = Var_t(\dbY(0))$. Therefore, unless $\pi_t = \pi$, we will have an additional term in the bias expression (note that $\beta_t \ne \beta$ in general, since $\beta_t$ is a function of time-varying factors). 

Finally, notice that in practice we must also estimate the potential outcomes absent treatment $Y_{it}(0)$, introducing another source of error that this derivation does not account for. Nevertheless, this exercise illustrates how controlling for the untreated potential outcomes absent treatment may reduce bias when the data generating process follows a linear factor model.
\end{remark}

\section{Other estimation strategies}\label{app:dr}

    While we focus on a regression-based estimation strategy, we briefly propose corresponding IPW and doubly-robust estimators. However, because we argue that a key advantage of our approach is its low variance relative to other methods, these approaches are less desirable in that they will, in general, lead to higher variance estimates than the regression estimator we discuss. In state policy settings, data are extremely limited so that a lower variance estimator is frequently worth the implicit bias-variance tradeoff. 
        
    First, we can obtain an IPW-estimator using the estimating equation,
    \begin{align*}
        \psi_1 = n^{-1}\sum_{i=1}^n Y_{it} \left[\frac{A_{it}}{e(Y_{i,t-1} - A_{i,t-1}\tau)} - \frac{1-A_{it}}{1-e(Y_{i,t-1} - A_{i,t-1}\tau)} \right] - \tau = 0.
    \end{align*}
    
    \noindent However, we cannot directly solve this without specification of $e(\cdot)$. Therefore, we may specify a parametric model $g\left\{e(Y_{i,t-1}; \tau)\right\} = \beta_0 + \beta_1\left(Y_{i,t-1} - A_{i,t-1}\tau\right)$ for a known link function $g(\cdot)$, and solve for these two sets of parameters jointly using M-estimation theory. Alternatively, we may combine the IPW and regression-based approach to construct a doubly-robust (DR) style estimator that solves the estimating equation $\psi_2$ below,
        {\small
        \begin{align}\label{eqn:drest}
        \psi_2 = n^{-1}\sum_{i=1}^n\left[\left(\frac{A_{it}}{e(Y_{i,t-1} - \tau A_{i,t-1})} - \frac{1-A_{it}}{1-e(Y_{i,t-1} - \tau A_{i,t-1})}\right)(Y_{it} - \mu_{A_{it}}(Y_{i,t-1}-\Pi A_{i,t-1})) + \Pi - \theta\right] = 0.
        \end{align}
        }
    \noindent where we again specify a parametric model for $e$ and where $\mu_A$ represents the L-DAM in \eqref{eqn:ldam}, and where we again solve for the parameter sets jointly using M-estimation. It is straightforward to modify both the IPW and DR estimators if Assumption \ref{asmpt-uniform} is implausible, or if time-specific effects are of particular interest. However, Assumption \ref{asmpt-uniform} connects the estimation approaches above to L-DAM specified in \eqref{eqn:ldam}. And again, in many studies of state policies, data are limited, making assumptions like Assumption \ref{asmpt-uniform} that reduce the number of parameters required for estimation worth the bias-variance trade-off.
        
    We briefly sketch below the result that under common regularity conditions, if either the propensity score model or the outcome model is correctly specified, the doubly-robust estimation will be consistent for the true causal effect. However, we caution that from a non-parametric perspective, the efficient influence function of $\theta$ depends on the density of $Y_{i,t-1}(0)$ and therefore has no non-parametric representation in terms of the observed data distribution. 

    \begin{proposition}\label{prop:dr}
    Let $e(Y_{i,t-1}-\tau A_{i,t-1}) = \P(A = 1\mid Y_{i,t-1}-\tau A_{i,t-1})$ and $\mu_a(Y_{i,t-1}-\theta A_{i,t-1}) = \E[Y_{it} \mid Y_{i,t-1} - \theta A_{i,t-1}, A_{it} = a]$. In equation (\ref{eqn:drest}), assume that $\hat e \pto \bar e$ and $\hat\mu_a \pto \bar\mu_a$ for $a = 0, 1$. Assume sufficient regularity conditions hold so that $\E[\hat e(Y_{i,t-1}-\hat\tau A_{i,t-1})] \pto \E[\bar e(Y_{i,t-1}-\tau A_{i,t-1})]$, and $\E[\hat \mu_a(Y_{i,t-1}-\hat\Pi A_{i,t-1})] \pto\E[\bar\mu_a(Y_{i,t-1}-\Pi A_{i,t-1})]$. Finally, assume that $\P(1 - \epsilon > \bar e(Y_{i,t-1} - \tau A_{i,t-1}) > \epsilon) = 1$ for $\epsilon > 0$. 
    
    If either $\bar e = e$ or $\bar\mu_a\pto\mu_a$ for $a=0,1$, then $\hat\theta \pto \theta$.
    \end{proposition}
    
    \begin{proof}[Proof sketch]\label{proof:dr}
    First, note that $\hat e \pto \bar e \implies \hat\tau \pto \tau$, and similarly $\hat\mu_a\pto\mu_a$ for $a=0,1$, implies $\hat\Pi\pto\Pi$. By the assumptions stated above and the law of iterated expectations,
    {\small
    \begin{align*}
    &\E[\hat\theta]\\
    &=\E\left[\left(\frac{A}{\hat e(Y_{i,t-1}-\hat\tau A_{i,t-1})} - \frac{1-A}{1-\hat e(Y_{i,t-1}-\hat\tau A_{i,t-1})}\right)(Y - \hat \mu_A(Y_{i,t-1}-\hat\Pi A_{i,t-1})) + \hat \Pi\right] \\
    &\pto \E\left[\left(\frac{A}{\bar e(Y_{i,t-1}-\tau A_{i,t-1})}(Y-\bar\mu_1(Y_{i,t-1}-\Pi A_{i,t-1})) - \frac{1-A}{1-\bar e(Y_{i,t-1}-\tau A_{i,t-1})}(Y -\bar\mu_0(Y_{i,t-1}-\Pi A_{i,t-1}))\right)+ \Pi\right] 
    \end{align*}
    }
    \noindent To ease notation, let $\bar e = \bar e(Y_{i,t-1} - \tau A_{i,t-1})$, $\bar \mu_a = \bar \mu_a(Y_{i,t-1}-\Pi A_{i,t-1})$ for $a = 0,1$, $e = e(Y_{i,t-1}(0))$, and $\mu_a = \mu_a(Y_{i,t-1}(0))$, where we again note by assumption that $Y_{i,t-1}(0) = Y_{i,t-1} - \theta A_{i,t-1}$. First, consider the case where $\bar e = e$. Then, 
    
    \begin{align*}
    \E\left[\frac{A(Y - \bar\mu_1)}{e}\right] &= \E\left[\frac{\E[AY \mid Y_{i,t-1}(0) ]}{e} - \frac{\E[A\bar\mu_1 \mid Y_{i,t-1}(0)]}{e}\right] \\
    &= \E\left[\frac{e(\mu_1-\E[\bar\mu_1 \mid A_{it},Y_{i,t-1}(0)])}{e}\right] \\
    &= \E[\mu_1 - \E[\bar \mu_1 \mid A_{it}, Y_{i,t-1}(0)]] \\
    &= \E[\mu_1 - \bar\mu_1^\star]
    \end{align*}
    
    \noindent Similar arguments hold for the other IPW term, so that we may conclude that
    
    \begin{align*}
    \E[\hat\theta] &\pto \E[(\mu_1 - \bar\mu_1^\star - (\mu_0 - \bar\mu_0^\star) + \bar\mu_1^\star-\bar\mu_0^\star] = \theta.
    \end{align*}
    
    \noindent Now consider the case where $\bar \mu_a = \mu_a$. Then we obtain, 
    
    \begin{align*}
    \E\left[\frac{A(Y - \mu_1)}{\bar e}\right] &= \E\left[\E\left[\frac{A(Y-\mu_1)}{\bar e} \mid Y_{i,t-1}(0), Y_{i,t-1}, A_{i,t-1}\right]\right]\\
    &=\E\left[\frac{1}{\bar e}\E\left(A(Y-\mu_1) \mid Y_{i,t-1}(0), Y_{i,t-1}, A_{i,t-1}\right)\right] \\
    &=\E\left[\frac{\P(A_{it} =1 \mid A_{i,t-1}, Y_{i,t-1}, Y_{i,t-1}(0))}{\bar e}\E\left(A(Y-\mu_1) \mid Y_{i,t-1}(0)\right)\right] \\
    &=\E\left[\frac{\P(A_{it} =1 \mid A_{i,t-1}, Y_{i,t-1}, Y_{i,t-1}(0))}{\bar e}(\mu_1-\mu_1)\right] = 0.
    \end{align*}
    
    \noindent where the first equality holds by iterating expectations, the second by the fact that $\bar e$ is fixed conditional on $Y_{i,t-1}(0), Y_{i,t-1}, A_{i,t-1}$ (since it is a deterministic function of the latter two elements), the third, fourth, and final equality by the fact that $\E[Y_{it} \mid A_{it}=1,A_{i,t-1}, Y_{i,t-1}, Y_{i,t-1}(0)] = \E[Y_{it} \mid A_{it} = 1, Y_{i,t-1}(0)]$, noting that $Y_{i,t-1}(0)= Y_{i,t-1} - \theta A_{i,t-1}$, and this expression is equal to $\mu_1$. A similar logic holds for the other IPW term and so that we conclude,
    
    \begin{align*}
    \E[\hat\theta] &\pto \theta.
    \end{align*}
    
    \end{proof}
    
    \end{document}